\documentclass[fleqn,usenatbib]{mnras}

\usepackage{newtxtext,newtxmath}
\usepackage[T1]{fontenc}

\DeclareRobustCommand{\VAN}[3]{#2}
\let\VANthebibliography\thebibliography
\def\thebibliography{\DeclareRobustCommand{\VAN}[3]{##3}\VANthebibliography}

\usepackage{graphicx}   % Including figure files

\usepackage{color,soul}  % Highlight changes following referee's report

\providecommand{\e}[1]{\ensuremath{\times 10^{#1}}}

\providecommand{\ion}[2]{\mbox{\textrm{#1\,\textsc{#2}}}}

\def\cm{\textrm{\thinspace cm}}

\def\erg{\textrm{\thinspace erg}}

\def\ga{\textrm{\thinspace gauss}}

\def\km{\textrm{\thinspace km}}

\def\Mpc{\textrm{\thinspace Mpc}}

\def\ps{\textrm{\thinspace s^{-1}}}

\def\s{\textrm{\thinspace s}}

\def\ergpscmps{\mbox{$\erg\cm^{-2}\s^{-1}\,$}}

\def\kmpspMpc{\mbox{$\km\ps\Mpc^{-1}$}}

\def\ps{\mbox{$\s^{-1}\,$}}

\def\la{\mbox{\textrm{L}$\alpha$}}

\def\ha{\mbox{\textrm{H}$\alpha$}}
\def\hb{\mbox{\textrm{H}$\beta$}}

\def\h0{\mbox{\textrm{H}$^0$}}
\DeclareMathAlphabet{\vib}{OML}{cmm}{m}{it}
\title[BASS XXXIX: Changing-look AGN]{BASS XXXIX: \textit{Swift}-BAT AGN with changing-look optical spectra}

\author[M. J. Temple et al.]{%
Matthew J. Temple,$^{1}$\thanks{E-mail: Matthew.Temple@mail.udp.cl}
Claudio Ricci,$^{1,2}$
Michael J. Koss,$^{3,4}$
Benny Trakhtenbrot,$^{5}$
Franz E. Bauer,$^{6,7,4}$
\newauthor
Richard Mushotzky,$^{8}$
Alejandra F. Rojas,$^{9,1}$
Turgay Caglar,$^{10}$
Fiona Harrison,$^{11}$
Kyuseok Oh,$^{12,13}$
\newauthor
Estefania Padilla Gonzalez,$^{14,15}$
Meredith C. Powell,$^{16}$
Federica Ricci,$^{17,18}$
Rog\'erio Riffel,$^{19}$
Daniel Stern,$^{20}$
\newauthor
and
C. Megan Urry$^{21}$
\\
$^{1}$N\'ucleo de Astronomía de la Facultad de Ingenier\'ia, Universidad Diego Portales, Av. Ej\'ercito Libertador 441, Santiago 22, Chile\\
$^{2}$Kavli Institute for Astronomy and Astrophysics, Peking University, Beijing 100871, People's Republic of China\\
$^{3}$Eureka Scientific, 2452 Delmer Street Suite 100, Oakland, CA 94602-3017, USA\\
$^{4}$Space Science Institute, 4750 Walnut Street, Suite 205, Boulder, Colorado 80301, USA\\
$^{5}$School of Physics and Astronomy, Tel Aviv University, Tel Aviv 69978, Israel\\
$^{6}$Instituto de Astrof\'{\i}sica  and Centro de Astroingenier{\'{\i}}a, Facultad de F\'{i}sica, Pontificia Universidad Cat\'{o}lica de Chile, Casilla 306, Santiago 22, Chile\\
$^{7}$Millennium Institute of Astrophysics (MAS), Nuncio Monse{\~{n}}or S{\'{o}}tero Sanz 100, Providencia, Santiago, Chile\\
$^{8}$Department of Astronomy, University of Maryland, College Park, MD 20742, USA\\
$^{9}$Centro de Astronom\'ia (CITEVA), Universidad de Antofagasta, Avenida Angamos 601, Antofagasta, Chile\\
$^{10}$Leiden Observatory, P.O. Box 9513, 2300 RA, Leiden, The Netherlands\\
$^{11}$Cahill Center for Astronomy and Astrophysics, California Institute of Technology, Pasadena, CA 91125, USA\\
$^{12}$Korea Astronomy and Space Science Institute, Daedeokdae-ro 776, Yuseong-gu, Daejeon 34055, Republic of Korea\\
$^{13}$Department of Astronomy, Kyoto University, Kitashirakawa-Oiwake-cho, Sakyo-ku, Kyoto 606-8502, Japan\\
$^{14}$Las Cumbres Observatory, 6740 Cortona Drive, Suite 102, Goleta, CA 93117-5575, USA\\
$^{15}$Department of Physics, University of California, Santa Barbara, CA 93106-9530, USA\\
$^{16}$Kavli Institute of Particle Astrophysics and Cosmology, Stanford University, 452 Lomita Mall, Stanford, CA 94305, USA\\
$^{17}$Dipartimento di Matematica e Fisica, Università Roma Tre, via della Vasca Navale 84, I-00146, Roma, Italy\\
$^{18}$INAF- Osservatorio di Astrofisica e Scienza dello Spazio di Bologna, via Gobetti 93/3, 40129 Bologna, Italy\\
$^{19}$Departamento de Astronomia, Instituto de F\'\i sica, Universidade Federal do Rio Grande do Sul, CP 15051, 91501-970, Porto Alegre, RS, Brazil\\
$^{20}$Jet Propulsion Laboratory, California Institute of Technology, 4800 Oak Grove Drive, MS 169-224, Pasadena, CA 91109, USA\\
$^{21}$Yale Center for Astronomy \& Astrophysics and Department of Physics, Yale University, P.O. Box 208120, New Haven, CT 06520-8120, USA
}

\date{Accepted 2022 November 08. Received 2022 November 02; in original form 2022 September 08}

\pubyear{2022}

\begin{document}
\label{firstpage}
\pagerange{\pageref{firstpage}--\pageref{lastpage}}
\maketitle

\begin{abstract}
%The abstract should briefly describe the aims, methods, and main results of the paper.
%It should be a single paragraph not more than 250 words (200 words for Letters).
%No references should appear in the abstract.
%
Changing-look (CL) AGN are unique probes of accretion onto supermassive black holes (SMBHs), especially when simultaneous observations in complementary wavebands allow investigations into the properties of their accretion flows.
We present the results of a search for CL behaviour in 412 \textit{Swift}-BAT detected AGN with multiple epochs of optical spectroscopy from the BAT AGN Spectroscopic Survey (BASS).
125 of these AGN also have 14--195\,keV ultra-hard X-ray light-curves from \textit{Swift}-BAT which are contemporaneous with the epochs of optical spectroscopy.
Eight CL events are presented for the first time, where the appearance or disappearance of broad Balmer line emission leads to a change in the observed Seyfert type classification.
Combining with known events from the literature, 21 AGN from BASS are now known to display CL behaviour. 
Nine CL events have 14--195\,keV data available, and five of these CL events can be associated with significant changes in their 14--195\,keV flux from BAT.
The ultra-hard X-ray flux is less affected by obscuration and so these changes in the 14--195\,keV band suggest that the majority of our CL events are
 not due to changes in line-of-sight obscuration.
We derive a CL rate of 0.7-6.2 per cent on 10-25 year time-scales, and show that many transitions happen within at most a few years. 
Our results motivate further multi-wavelength observations with higher cadence to better understand the variability physics of accretion onto SMBHs.
% 231 words
\end{abstract}


%%%%%%%%%%%%%%%%%%%%%%%%%%%%%%%%%%%%%%%%%%%%%%%%%%

%%%%%%%%%%%%%%%%% BODY OF PAPER %%%%%%%%%%%%%%%%%%

\section{Introduction}
%This is a simple template for authors to write new MNRAS papers.
%See \texttt{mnras\_sample.tex} for a more complex example, and \texttt{mnras\_guide.tex} for a full user guide.
%All papers should start with an Introduction section, which sets the work
%in context, cites relevant earlier studies in the field
%and describes the problem the authors aim to solve.

In the simplest unified model for active galactic nuclei (AGN), orientation determines their observed properties \citep[][]{1993ARA&A..31..473A, 1995PASP..107..803U}. 
When the obscuring toroidal material is blocking the line of sight, soft X-ray, ultraviolet and optical emission from the nucleus is absorbed, as is light from the broad emission line region (BLR); only narrow line emission is observed in the ultraviolet and optical. 
While the unified model can explain many observed properties of AGN, a growing number of highly variable sources challenge the simplest version of this model.
Such `changing-look' (CL) AGN experience rapid changes in their optical, ultraviolet or X-ray classification. 
In the rest-frame optical and ultraviolet,  AGN have been seen to rapidly change between different Seyfert spectral type classifications: 
 from a type 1 source with strong, broad  emission lines to a type 2 source with only narrow emission lines or vice versa \citep[][]{1973A&A....22..343C, 2014ApJ...788...48S, 2015ApJ...800..144L, 2016ApJ...826..188R, 2016ApJ...821...33R, 2018ApJ...862..109Y, 2019ApJ...874....8M, 2020MNRAS.491.4925G, 2020ApJ...905...52G, 2021MNRAS.503.2583S, 2022MNRAS.513L..57L}.
In multi-epoch X-ray observations, AGN have been seen to change from Compton-thick to Compton-thin or vice versa, consistent with a large change in the amount of material which is obscuring the line-of-sight to the continuum source \citep[e.g.,][]{2002MNRAS.329L..13G, 2003MNRAS.342..422M, 2012MNRAS.421.1803M, 2016ApJ...820....5R}. 

However, changes in line-of-sight obscuration can only account for a subset of CL AGN. 
In some of these objects, known as `changing-state' (CS) AGN, changes in the structure of the accretion disc are required to explain the characteristics of the observed variability.
For example, \citet{2018ApJ...864...27S} showed that the mid-infrared emission was highly variable in the CS quasar WISE\,J105203.55+151929.5 on time-scales of only a few years - significantly shorter than would be expected for an obscurer to cover the mid-infrared emitting region.
While the geometry of the accretion flow will vary with large changes in the accretion rate  \citep{1995ApJ...452..710N, 2014ARA&A..52..529Y, 2019A&A...630A..94G, 2019ApJ...883...76R}, 
such changes are expected to occur on the viscous time-scale, which is tens to thousands of years in the discs around supermassive black holes (SMBHs).
The changes observed in CS AGN often occur within tens of months, which is closer to the thermal time-scale \citep[e.g.][]{2017ApJ...835..144G}, suggesting that this behaviour is driven by temperature variations or instabilities in the inner accretion disc
\citep{2018MNRAS.480.3898N, 2018MNRAS.480.4468R}.
CS\,AGN are thus an important probe of accretion disc physics, especially when coupled with multi-wavelength observations in the rest-frame X-ray, ultraviolet, optical and infrared which probe the different line- and continuum-emitting regions around the SMBH.

Over the past years, more than one hundred CL AGN have been discovered at various redshifts, but only
a few of these objects have been studied in detail in the X-ray band. 
Most famously, the appearance and disappearance of broad optical lines in Mrk\,1018 \citep{1986ApJ...311..135C, 2016A&A...593L...8M} have been associated with the evolution of the X-ray--ultraviolet continuum emission which is responsible for photoionizing the BLR gas \citep{2018MNRAS.480.3898N, 2021MNRAS.506.4188L, 2022ApJ...930...46L}.  
More dramatic behaviour was observed in 1ES\,1927+654, which underwent a complete transformation of its X-ray spectral properties, and showed X-ray variability of over four orders of magnitude on time-scales of months \citep{2019ApJ...883...94T, 2020ApJ...898L...1R, 2021ApJS..255....7R}.
However, due to the relatively small number of CS AGN observed in the X-rays, it is  still unclear whether these events are always accompanied by a clear transition in the X-ray band. 
It is therefore critical to identify more examples of AGN which both show changes in their optical spectral type classification and also have contemporaneous multi-wavelength data which can be used to constrain the physics of CL and CS AGN transitions.

At the same time, we are moving from serendipitous discoveries to an era of systematically searching for CL behaviour in large surveys. Most CL AGN studies to date have  looked for changes in the emission properties of a parent sample which has been selected using optical data. 
Such searches tend to be biased towards finding either the appearance of broad lines in previously known type 2 objects (so-called `turn-on' events), or the disappearance of broad emission features from previous type 1 objects (`turn-off' events).
For example, recent work from the SDSS-IV TDSS survey found 15 turn-off and 4 turn-on CL AGN, starting from a sample of 64\,039 broad-line SDSS quasars \citep{2022ApJ...933..180G}.
A complementary search by \citet{2021MNRAS.tmp.3368H} found 24 turn-on and 4 turn-off CL AGN, starting from a sample of 1092 type 2 and 304 type 1 AGN in the 6dF Galaxy Survey.
To better understand the incidence of CL, and specifically CS behaviour, it is therefore desirable to search within a parent sample that was not selected based on ultraviolet or optical properties, thus being less biased towards either of the main spectral AGN sub-classes (type 1 or type 2 AGN).

The Burst Alert Telescope \citep[BAT;][]{2005SSRv..120..143B} onboard the \textit{Swift} satellite \citep{2004ApJ...611.1005G} has been scanning the sky in the 14--195\,keV X-ray band since December 2004.
The BAT AGN Spectroscopic Survey %\footnote{\url{www.bass-survey.com}}
(BASS; \citealt{Koss_DR2_overview}) is a comprehensive, multi-wavelength effort to
investigate the properties of local AGN selected via their \textit{Swift}-BAT 14--195\,keV X-ray emission. 
This ultra-hard X-ray selection is far less biased by obscuration \citep[fig.\,1 of][]{2015ApJ...815L..13R}, effectively ensuring a complete census of actively accreting SMBHs in the local universe.
The second data release (DR2) from BASS focuses on the 858 AGN \citep{Koss_DR2_catalog} from the 70-month \textit{Swift}-BAT catalogue \cite[2004 December to 2010 October;][]{2013ApJS..207...19B}. 
BASS DR2 includes 1449 optical spectra for these 858 objects, from which have been derived emission line measurements, black hole masses, and accretion rates \citep{denBrok_DR2_NIR, Koss_DR2_catalog, Koss_DR2_sigs, Mejia_Broadlines, Oh_DR2_NLR, Ricci_DR2_NIR_Mbh, 2022ApJ...938...67R}.
Due to their bright fluxes, BASS AGN are some of the most well-studied AGN in the universe, with detailed observations across all wavebands in many cases. This sample therefore provides an excellent test bed for constraining the physics of CL AGN.

In this paper, we present a sample of eight new CL events identified through visual inspection of 412 AGN with $\geq$2 epochs of optical spectroscopy in the BASS DR1 and DR2 catalogues.
We also discuss the ultra-hard X-ray light-curves of five previously known CL AGN from the BASS sample; four from the 70-month BAT catalogue which formed the basis for BASS DR2, and one additional object from the 105-month BAT catalogue \citep{2018ApJS..235....4O}.
In Section~\ref{data}, we describe the data set and the methods used to identify the sample of CL AGN, which we present in Section~\ref{sec:sample}. These objects are discussed and compared to previously known populations in Section~\ref{sec:discuss}.
Throughout this work we assume a flat concordance cosmology with $\Omega_\Lambda=0.7, \Omega_m=0.3, H_0=70\,\kmpspMpc$, consistent with previous papers from the BASS collaboration.

\section{Data}
\label{data}

\begin{figure}
    \centering
    \includegraphics[width=\columnwidth]{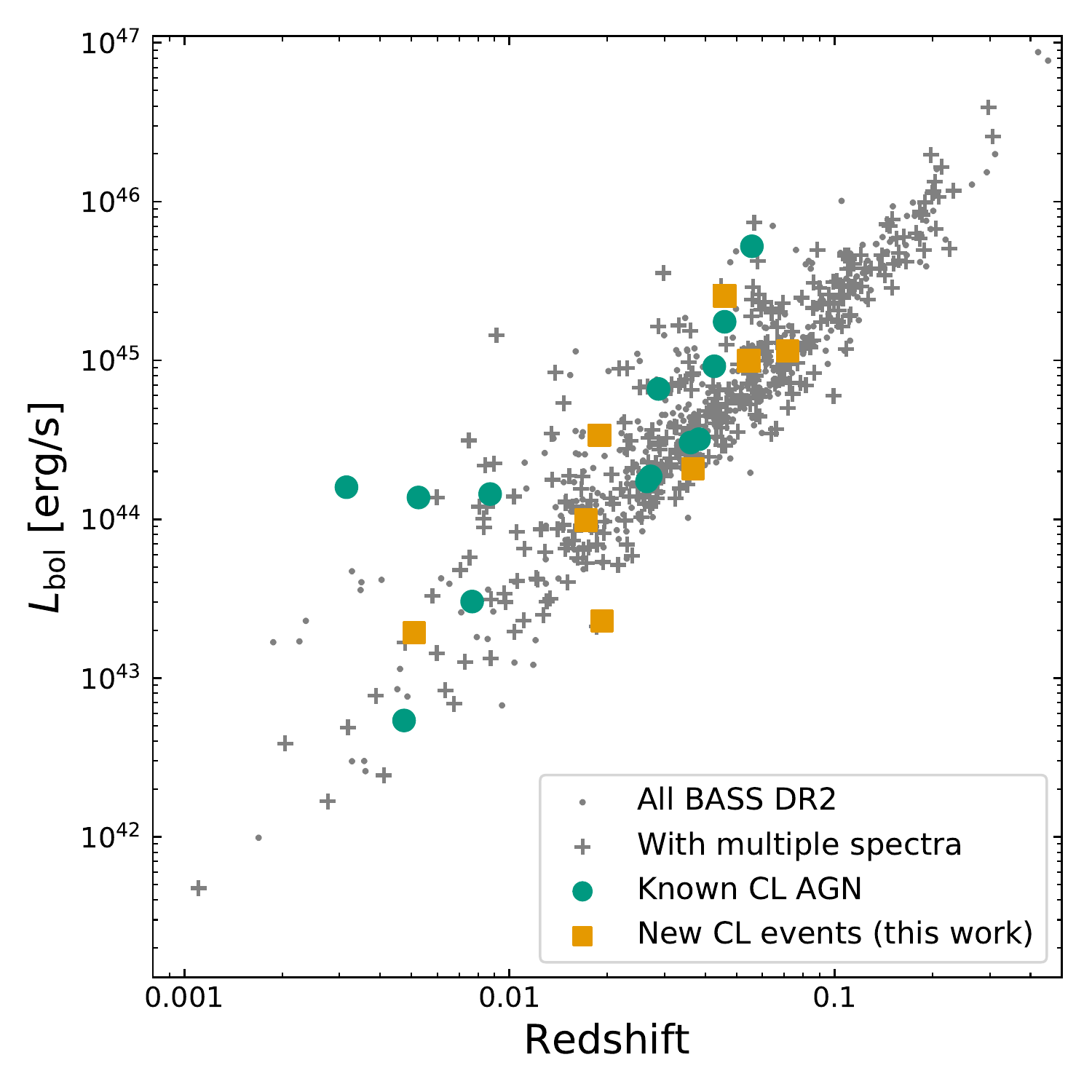}
    \caption{Distribution in redshift--luminosity space for BASS DR2, with the parent sample for this investigation marked with crosses. Previously known and newly identified CL AGN (see Section~\ref{sec:sample}) are identified with coloured symbols.
    We do not see any changes in type in BASS AGN with redshifts $z>0.1$, which are among the most luminous objects in our parent sample. }
    \label{fig:zL}
\end{figure}

\subsection{Optical spectra}
\label{sec:data}

We start from a sample of 2168 unique optical spectra covering 1105 unique BAT-detected AGN, consisting of the complete BASS DR1 and DR2 catalogs \citep{2017ApJ...850...74K, Koss_DR2_catalog} as well as additional, unpublished observations obtained as part of the ongoing BASS effort to study newly identified AGN in the BAT 105-month survey. Blazars \citep[][]{2019ApJ...881..154P, 2022arXiv220909929M} and other types of beamed AGN were then removed, using the classifications reported in the \citet{Koss_DR2_catalog} catalogue. High spectral resolution observations with limited wavelength coverage, typically only $\approx$8000--9000\AA, were also discarded. Such spectra were taken to measure stellar velocity dispersions via the Calcium triplet absorption complex and do not cover the broad Balmer emission lines.
From the remaining data, 430 BAT sources have multiple epochs of optical spectroscopy.
The merging system Was\,49 (BAT ID 605) has one spectrum from each of its two nuclei in the BASS DR2 catalogue and is removed from our sample.
17 sources have redshifts $z>0.5$, where their observed-frame optical spectra cover the rest-frame ultraviolet regime. Such objects were inspected separately, and no significant changes were observed in their rest-ultraviolet emission features.

\begin{figure*}
    \centering
    \includegraphics[width=\columnwidth]{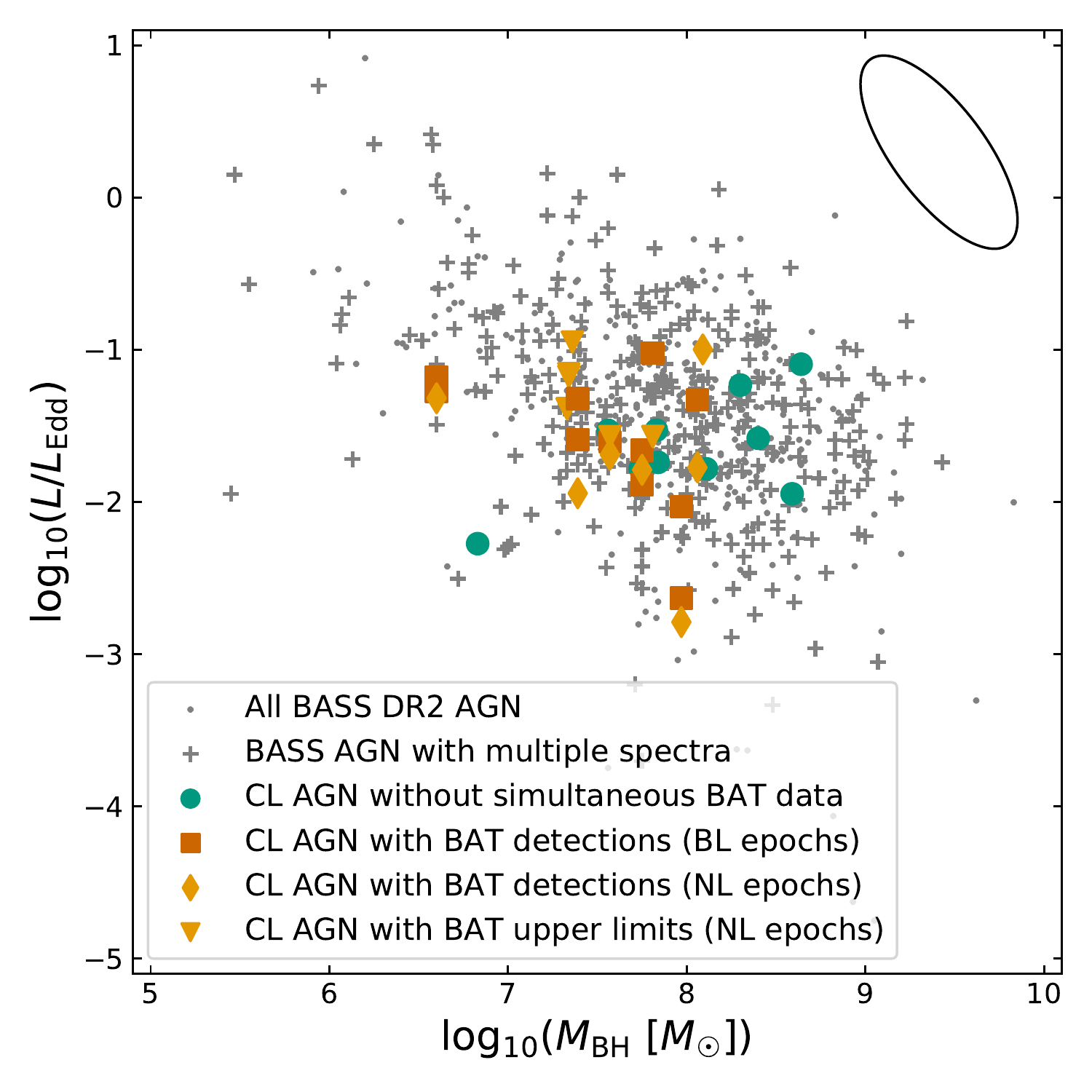}
    \includegraphics[width=\columnwidth]{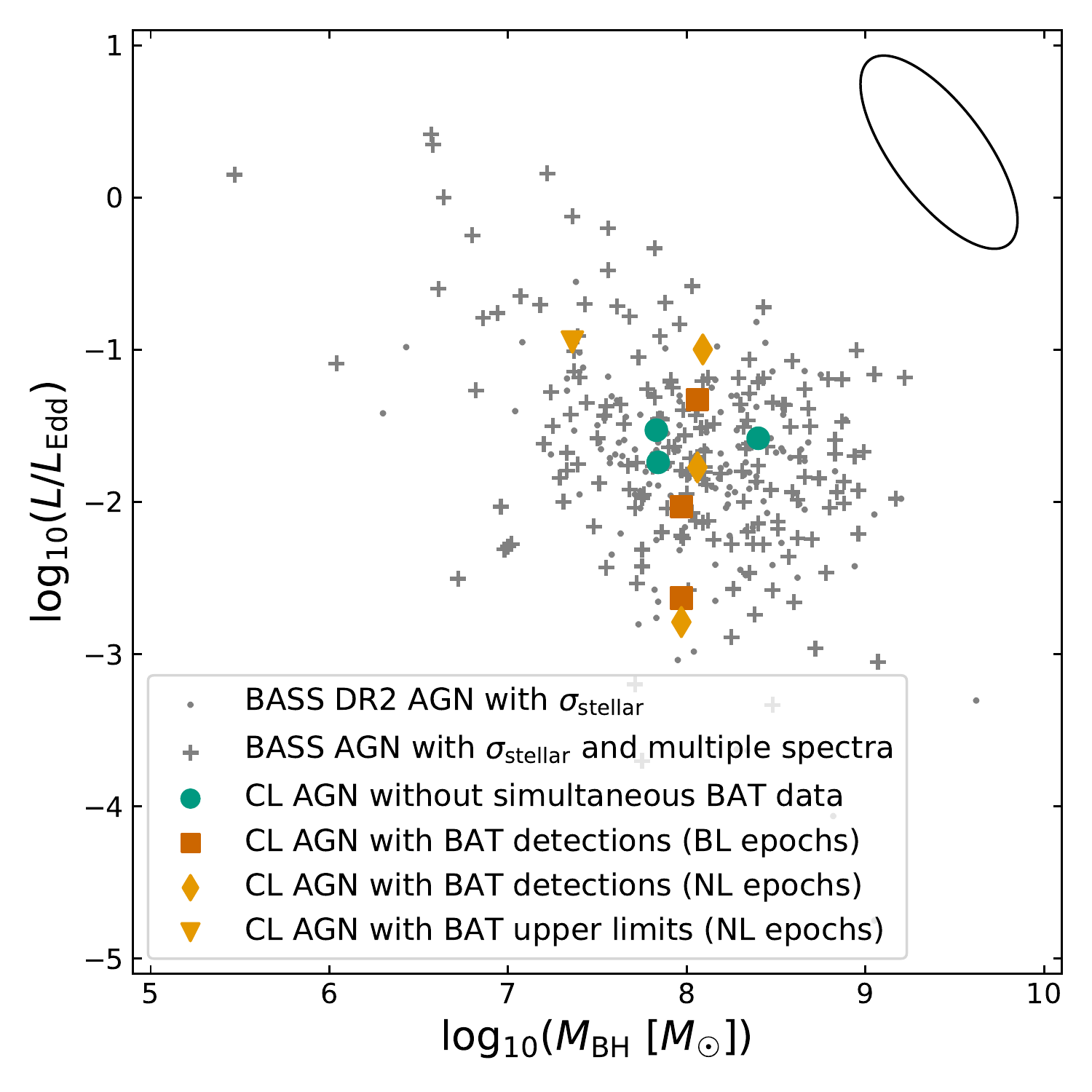}
    \caption{
    \textit{Left:} 747 unbeamed $z<0.5$ AGN  with black hole mass and Eddington ratio estimates  from BASS DR2 \citep{Koss_DR2_catalog}.
     Bolometric luminosities are estimated using  $L_\textrm{bol}=15\times L_\textrm{intrinsic 2-10\,keV}$ where available from \citet{2017ApJS..233...17R}, else $L_\textrm{bol}=8\times L_\textrm{14-195\,keV}$.
     \textit{Right:} as left, but only including objects with $M_\textrm{BH}$ inferred from measurements of the host galaxy stellar velocity dispersion \citep{Koss_DR2_sigs}.
     Crosses show objects with multiple epochs of optical spectroscopy which form the parent sample for this work. 
     CL AGN are shown in orange where $L/L_\textrm{Edd}$ can be inferred from simultaneous BAT light-curves (Section~\ref{sec:mbh}) with different shapes representing epochs where broad Balmer lines are seen (BL epochs) and epochs where only narrow lines are observed (NL epochs).
    Green circles show CL AGN where the change in state is not covered by the 157-month \textit{Swift}-BAT survey, and $L/L_\textrm{Edd}$ is instead taken from \citet{2017ApJS..233...17R}.
    The ellipse shows the 1$\sigma$ covariant uncertainty arising from  representative 0.45\,dex uncertainties on both $M_\textrm{BH}$ and $L_\textrm{bol}$ \citep{Koss_DR2_catalog}.
    %Our CL AGN have $M_\textrm{BH}$ estimates in the range $10^{6-9}M_\odot$. 
    All our CL AGN appear to have $L/L_\textrm{Edd} \lesssim 0.1$, as expected if CL behaviour arises from changes in the accretion disc structure \citep{2018MNRAS.480.3898N, 2019ApJ...883...76R}.
    %{Vermillion RGB=0.80,0.40,0 or orange RGB=0.90,0.60,0 and blueish green RGB=0,0.60,0.50}
    }
    \label{fig:BHMs}
\end{figure*}

Following these cuts, we have 412 AGN from BASS DR2 with multiple epochs of rest-frame optical spectroscopy, which form the parent sample for this investigation. Within this parent sample, the median number of spectral epochs per object is 3. These sources cover the full range of $L_\textrm{bol} (\approx 10^{42-47}\,\textrm{erg}\,\textrm{s}^{-1})$, $M_\textrm{BH} (\approx 10^{6-9.5}\,M_\odot)$ and $L/L_\textrm{Edd} (\approx 10^{-4}-10^1)$ of the BASS DR2 population, as shown in Figs.~\ref{fig:zL} and \ref{fig:BHMs}. 

Each of these 412 AGN were visually inspected
for changes in their Balmer line (\hb\ and/or \ha) properties. 
Due to the heterogeneity of the spectroscopic data for our sample, acquired using 24 different instruments and 18 different telescopes, the accuracy of the spectrophotometry (i.e.\ the absolute flux calibration) of each spectrum varies considerably.
Throughout this work we therefore normalise the flux in each spectral band (i.e.\ around \hb\ and \ha) by dividing out the median flux, to better visualise the changes in the emission lines.
Any change in the seeing or the aperture size (slit width or fibre diameter) will lead to a change in the relative contribution from the host galaxy to the flux of the extracted spectrum, and hence a change in the equivalent width of any nuclear emission features.
However, each spectrum in our parent sample covers the nucleus of the galaxy and so if broad line emission is present then we expect to detect it.
The aim of this work is to find high-confidence CL AGN in BASS spectra, and to analyse those which could be due to CS behaviour, but not to accurately quantify their emission line properties.
To mitigate the effects of varying signal-to-noise ratio, spectral resolution, instrument and flux calibration, herein we present only objects which were seen with complete changes in spectral type, i.e.\ the complete appearance or disappearance of broad \hb\ or \ha\ emission lines (Fig.~\ref{fig:specs}).
In Appendix~\ref{sec:app} we show examples of spectra which did not meet this criteria and so were discarded during our visual inspection process.

\subsection{BAT light-curves}
\label{sec:BAT_LCs}

While the AGN in BASS DR2 were selected from the first 70 months of the BAT survey, data is now available covering the first 157 months of \textit{Swift} operations, from 2004 December to 2017 December (Lien et al.\ in prep.).
We rebin each light-curve to 6-month bins to improve the signal-to-noise ratio and to aid the identification of longer-term trends in the 14--195\,keV emission.

From our parent sample of 412 AGN, 162 spectral epoch-pairs from 125 objects occur within the period 2004 December to 2017 December. The vast majority of the remaining 287 objects have only one spectrum from before 2017 December with subsequent spectral epochs (usually from targeted BASS programs) dating from 2018 January onwards.  We therefore have X-ray light-curves contemporaneous with multi-epoch optical spectra in 125 AGN, making this the largest search for CL AGN to date from simultaneous optical and X-ray multi-epoch data.

\subsection{Black hole masses and Eddington ratios}
\label{sec:mbh}
We use the black hole mass ($M_\textrm{BH}$) estimates  from the BASS DR2 catalogue. These measurements are described in detail by \citet{Koss_DR2_catalog}, and here we only briefly discuss these estimates. $M_\textrm{BH}$ estimates are taken from (i) literature measurements of spatially resolved megamasers, stellar or gas dynamics, or reverberation mapping campaigns; (ii) single-epoch measurements of broad emission lines \citep{Mejia_Broadlines}; and (iii) measurements of stellar velocity dispersions \citep{Koss_DR2_sigs} and assuming the \citet{2013ARA&A..51..511K} $M_\textrm{BH}-\sigma_\ast$ relation.

Due to the variable nature of the CL AGN studied in this work, the question naturally arises as to whether the BLR in these objects is virialised and whether the usual virial scaling relations used for broad emission lines hold true. 
In the right hand panel of Fig.~\ref{fig:BHMs} we show the subset of the catalogue with  $\sigma_\ast$-derived $M_\textrm{BH}$.
As would be expected, these sources do not include many of the brightest AGN inferred to have $L/L_\textrm{Edd} \gtrsim 1$, but do cover the full range of parameter space spanned by the CL AGN in this work.
Recently \citet{2021arXiv211207284J} showed that the virial $M_\textrm{BH}$ estimated from the bright epochs in a sample of 26 CL AGN is consistent with the $M_\textrm{BH}$ inferred from measurements of stellar velocity dispersions in their faint epochs, suggesting both that (i) CL AGN follow the usual $M_\textrm{BH}-\sigma_\ast$ relation and (ii) the virial scaling relations used to derive $M_\textrm{BH}$ are still applicable in CL AGN. 
The dominant uncertainty on each $M_\textrm{BH}$ estimate is of order 0.45\,dex  due to the systematic uncertainties in virial and $\sigma_\ast$ scaling relations \citep{2013BASI...41...61S, Koss_DR2_catalog}. This gives rise to the covariant uncertainty ellipse in the $M_\textrm{BH}$--$L/L_\textrm{Edd}$ space  shown in Fig.~\ref{fig:BHMs}.

For the time-variable sources studied in this work, any estimation of the Eddington ratio also requires  knowledge of the instantaneous bolometric luminosity $L_\textrm{bol}$.
For spectra taken between 2005 and 2017 (inclusive) we infer an estimate of $L_\textrm{bol}$ using the relevant \textit{Swift}-BAT 14--195\,keV light-curve, assuming 1 Crab is 2.3343\e{-8}\ergpscmps \citep{2018ApJS..235....4O}, a $\Gamma=1.8$ continuum power-law index \citep{2017ApJS..233...17R}, and a bolometric correction consistent with previous BASS works \citep{2009MNRAS.392.1124V, Koss_DR2_catalog}:
\begin{equation}
    L_\textrm{bol} = 8 \times L_{14-195} = 8 \times 4\pi\textrm{d}_\textrm{L}^2\frac{F_{14-195}}{(1+z)^{2-\Gamma}},
\end{equation}
where $F_{14-195}$ is the weighted-average 14--195\,keV flux from BAT in the 6 months prior to the date of the relevant optical spectrum.

\section{Results}
\label{sec:sample}

CL AGN can be divided into two broad categories \citep[e.g.,][]{2022arXiv221105132R}: those due to changes in line-of-sight obscuration, and those due to intrinsic variability such as changes in the structure of the accretion flow or the BLR.
Here we discuss each CL AGN individually, with the aim of distinguishing between those which could be ascribed to obscuration, and those which are \textit{bona fide} CS AGN. The ultra-hard X-ray flux measured by  \textit{Swift}-BAT is not only probing the power emitted by the accretion disc, but also remains unaffected by changes in obscuration on the level of $N_\textrm{H}\lesssim10^{24}$\,cm$^{-2}$ \citep{2015ApJ...815L..13R,2016ApJ...825...85K}, meaning that any change in the BAT flux contemporaneous with an optical CL event is most likely due to CS transition.
We therefore include discussion of the BAT light-curves where available for each CL event.

\subsection{New CL events}
\label{sec:new_CSAGN}

\begin{figure*}
    \centering
    \includegraphics[width=\columnwidth, clip=on, trim={0 15 0 0}]{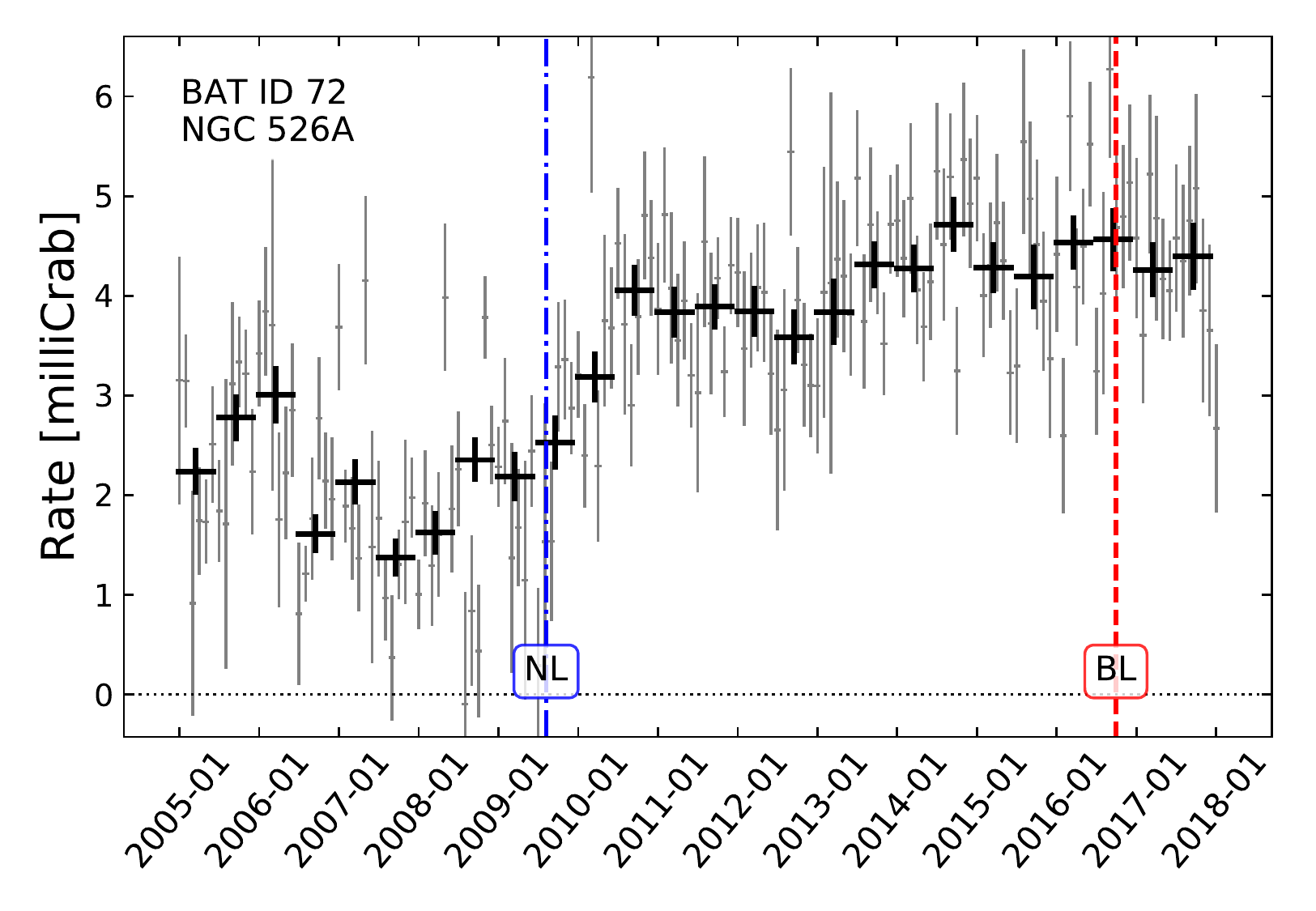}
    \includegraphics[width=\columnwidth, clip=on, trim={0 15 0 0}]{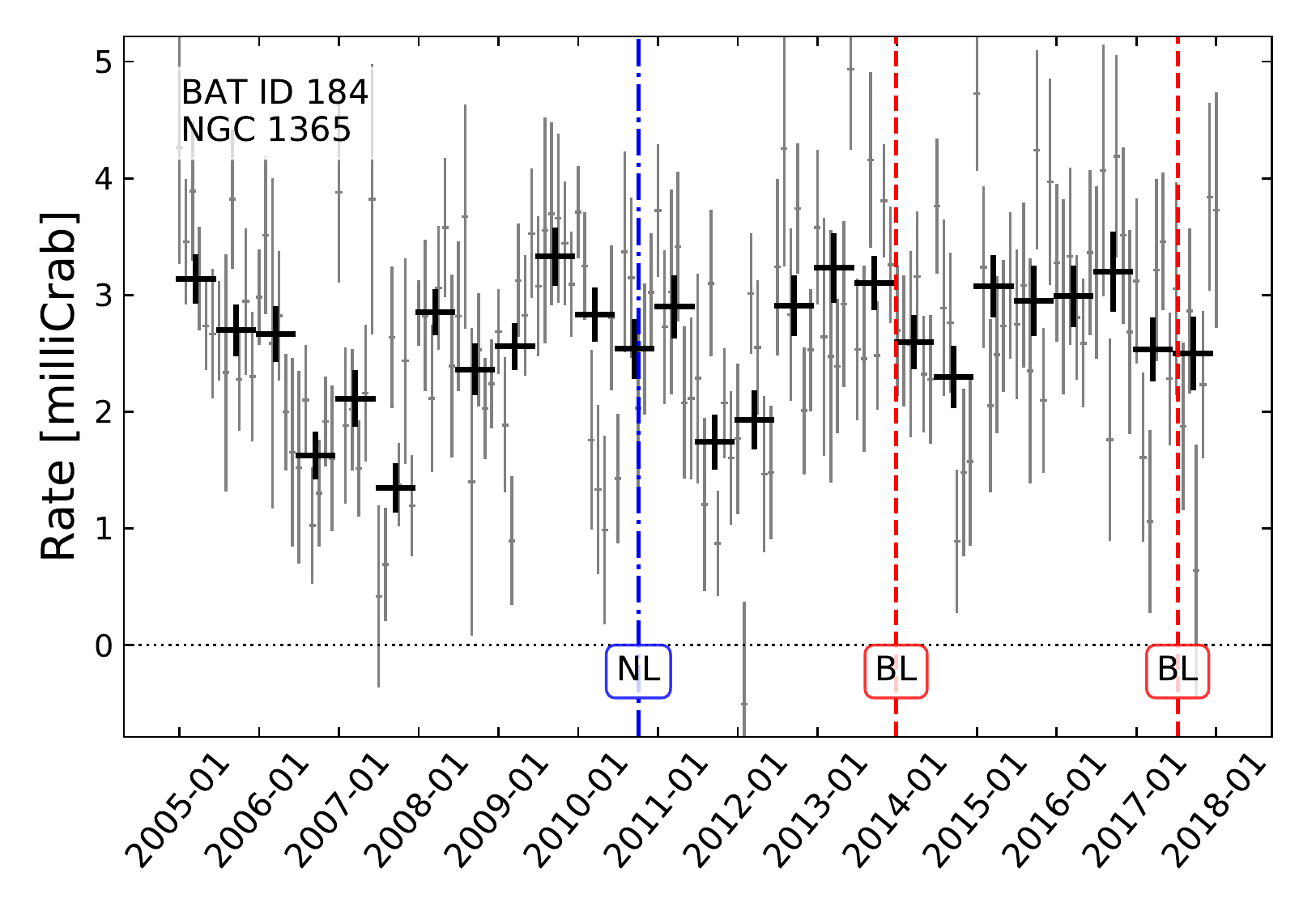}
    \includegraphics[width=\columnwidth, clip=on, trim={0 15 0 0}]{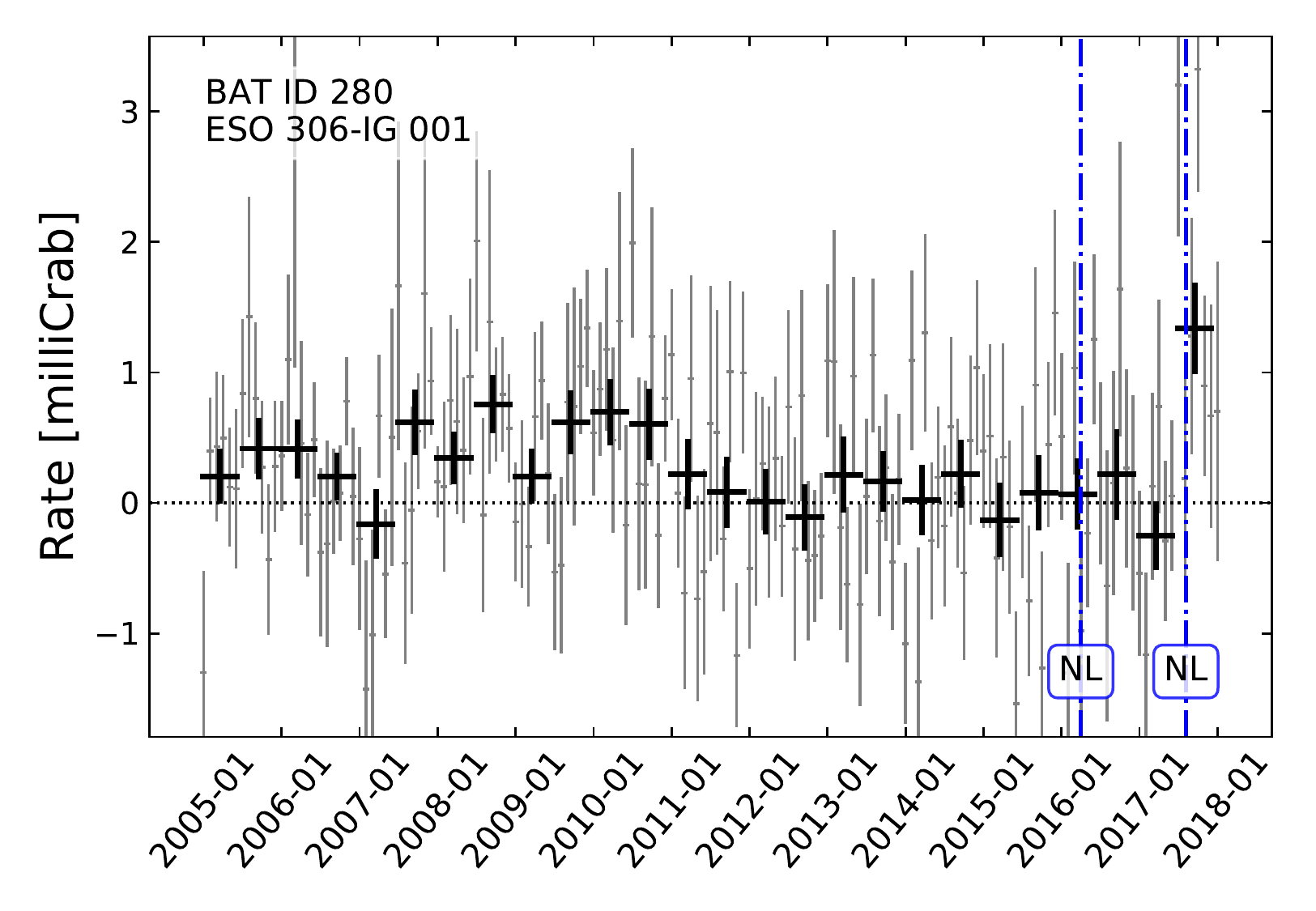}
    \includegraphics[width=\columnwidth, clip=on, trim={0 15 0 0}]{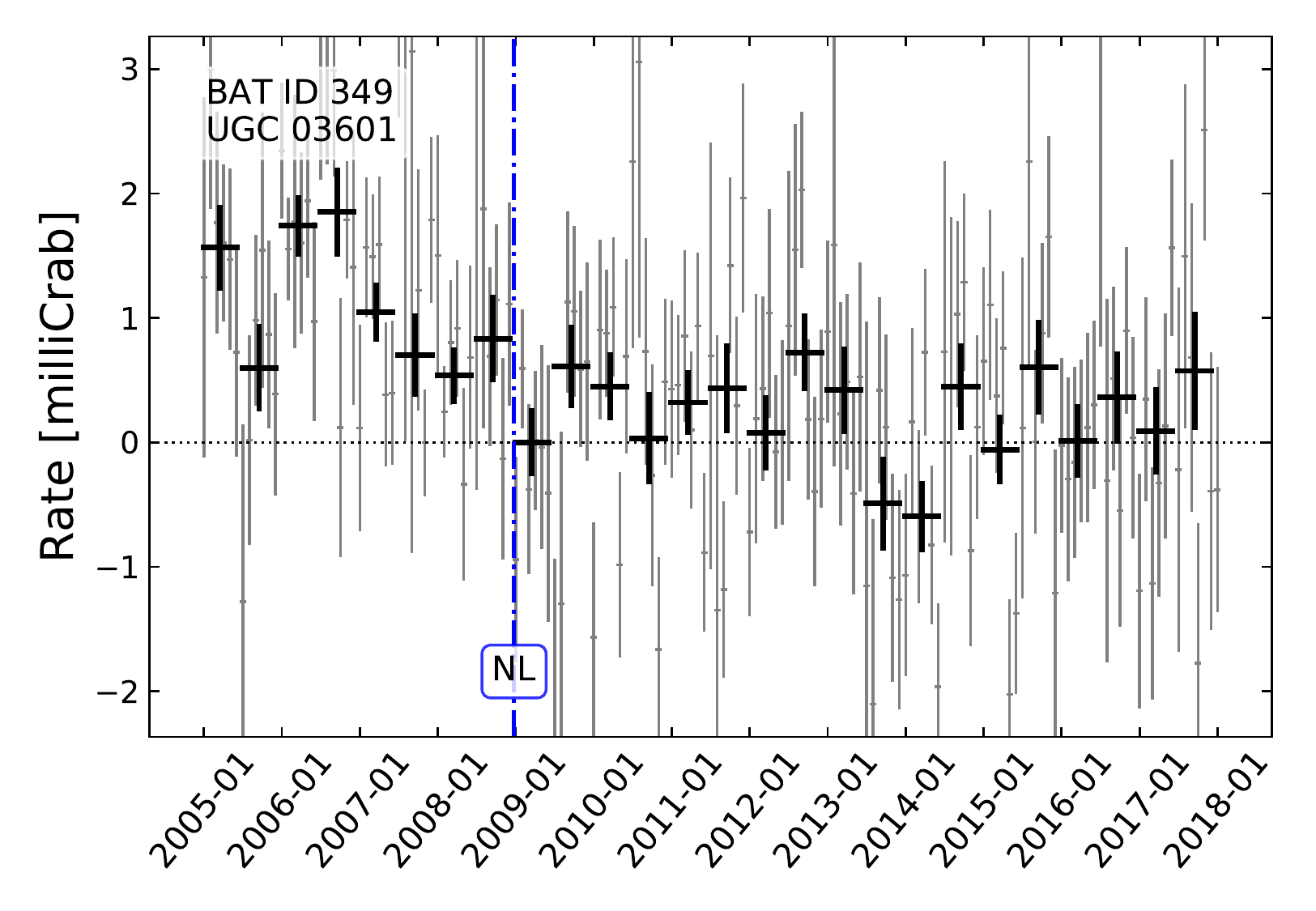}
    \includegraphics[width=\columnwidth, clip=on, trim={0 15 0 0}]{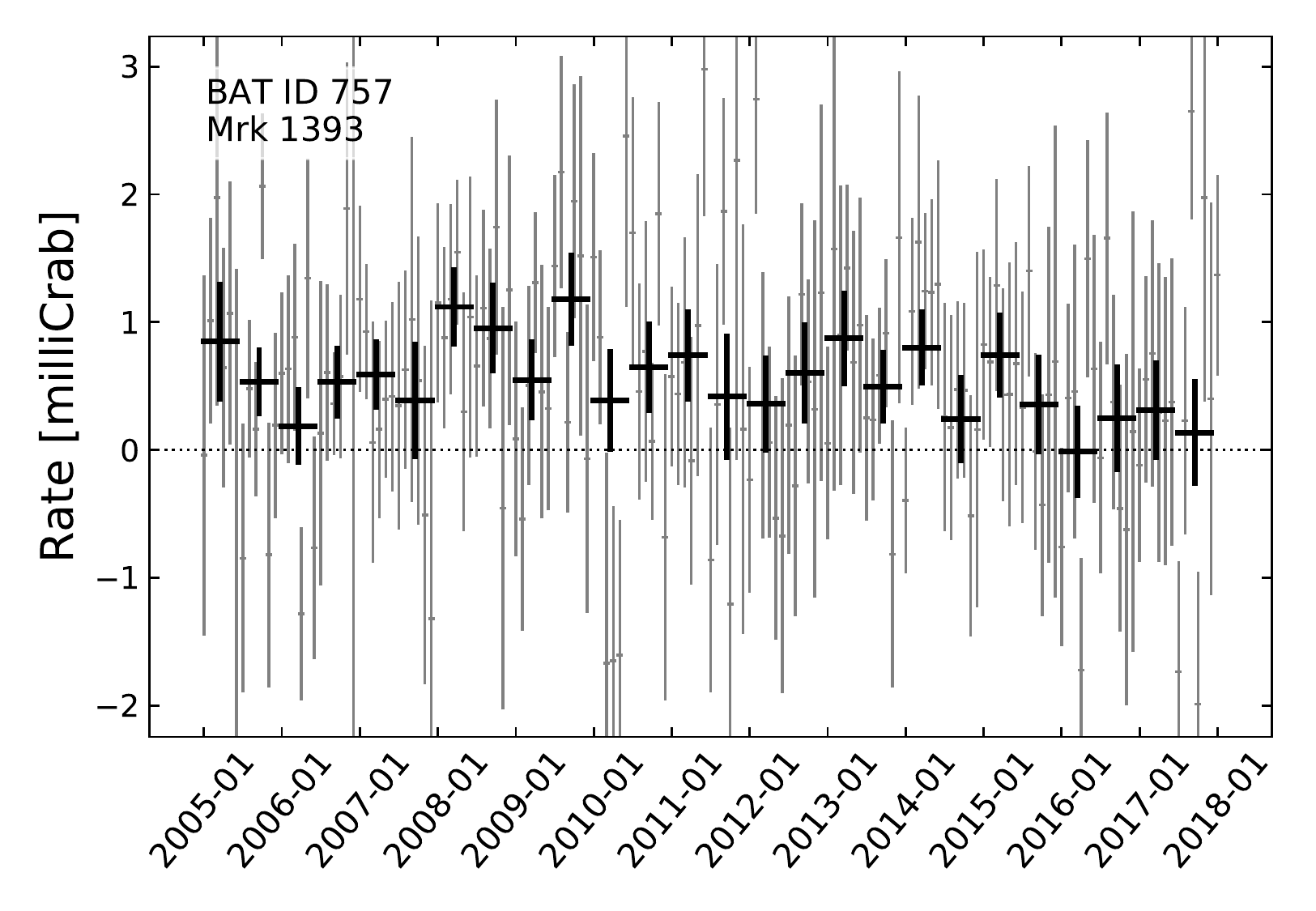}
    \includegraphics[width=\columnwidth, clip=on, trim={0 15 0 0}]{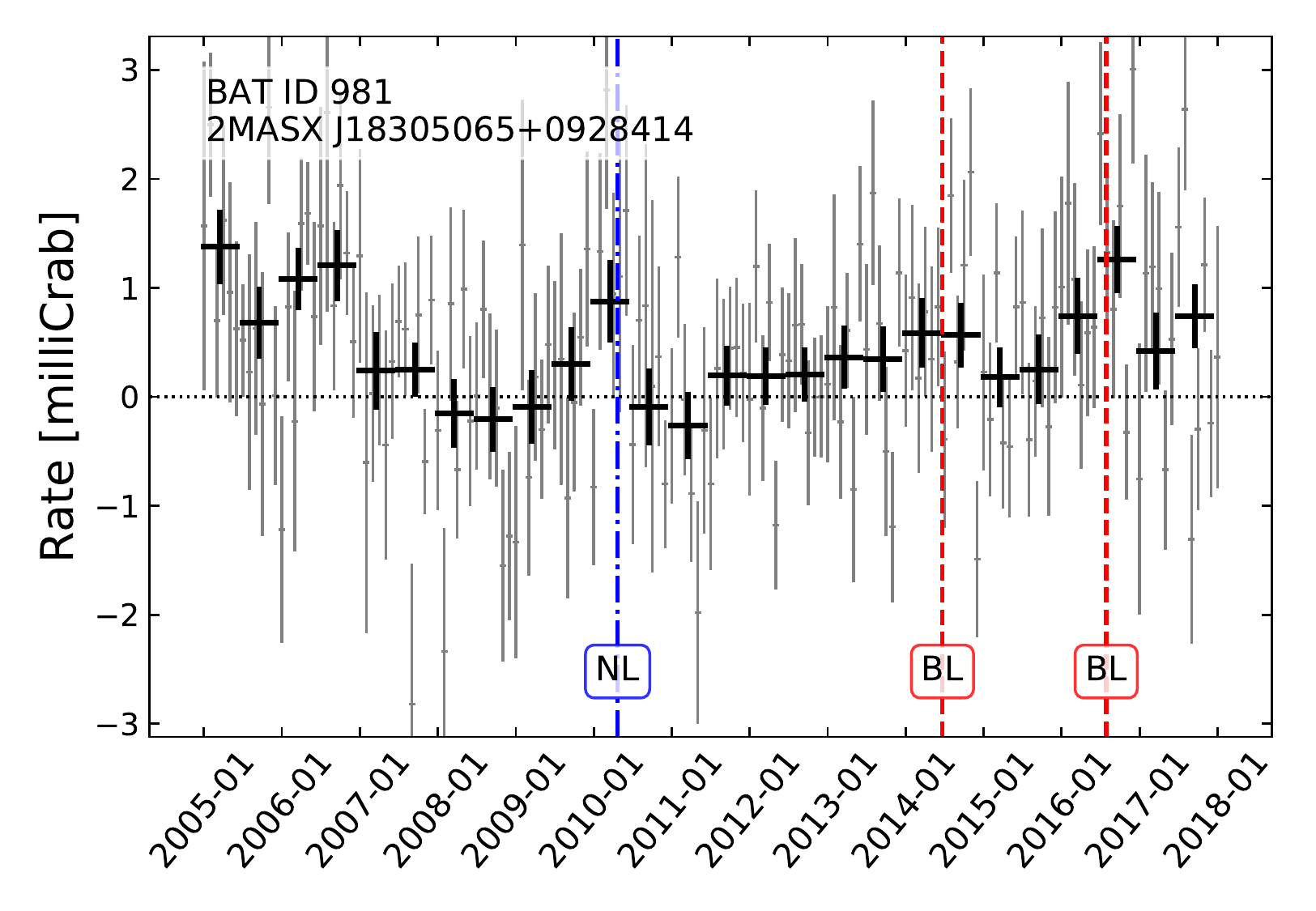}
    \includegraphics[width=\columnwidth, clip=on, trim={0 15 0 0}]{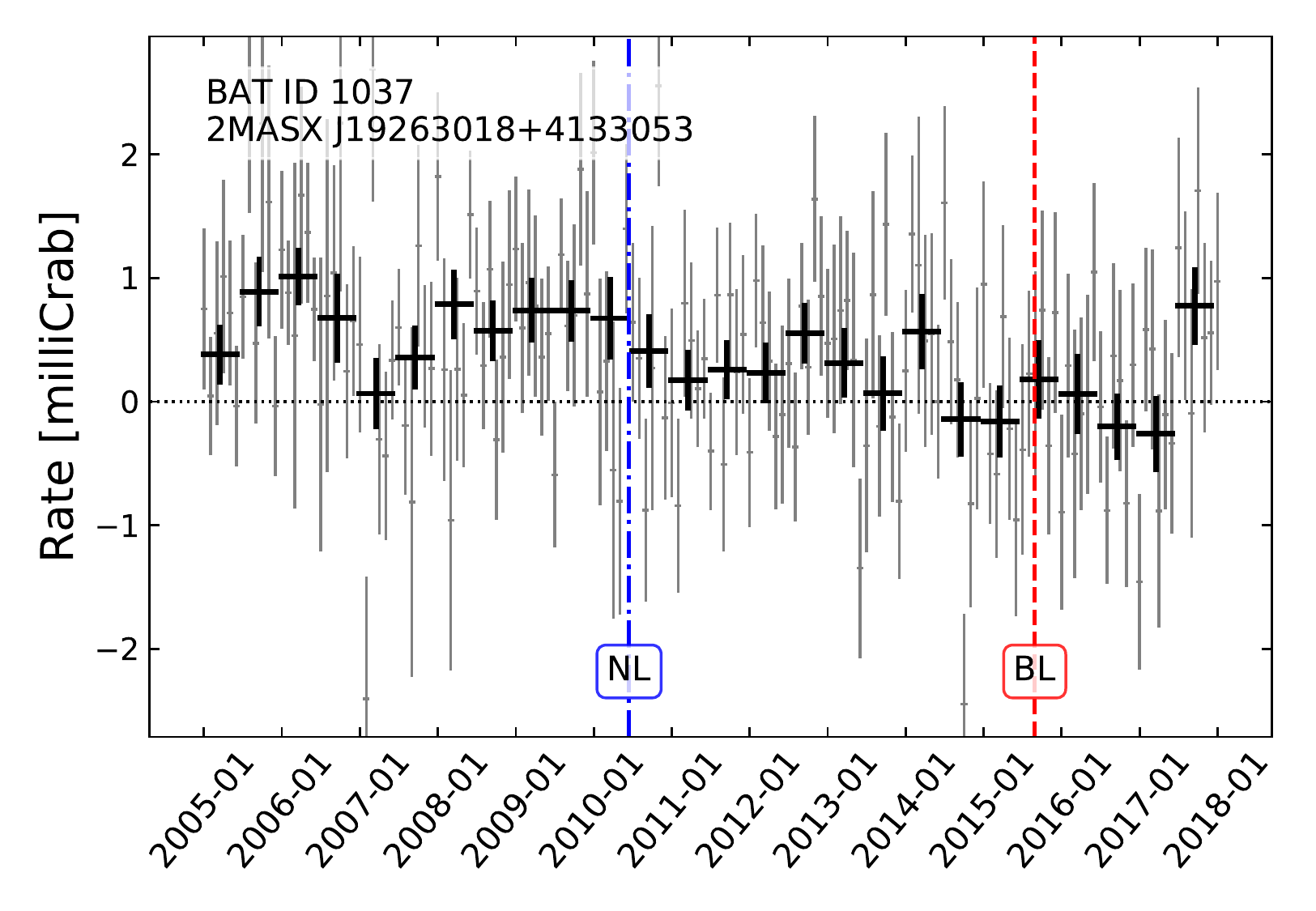}
    \includegraphics[width=\columnwidth, clip=on, trim={0 15 0 0}]{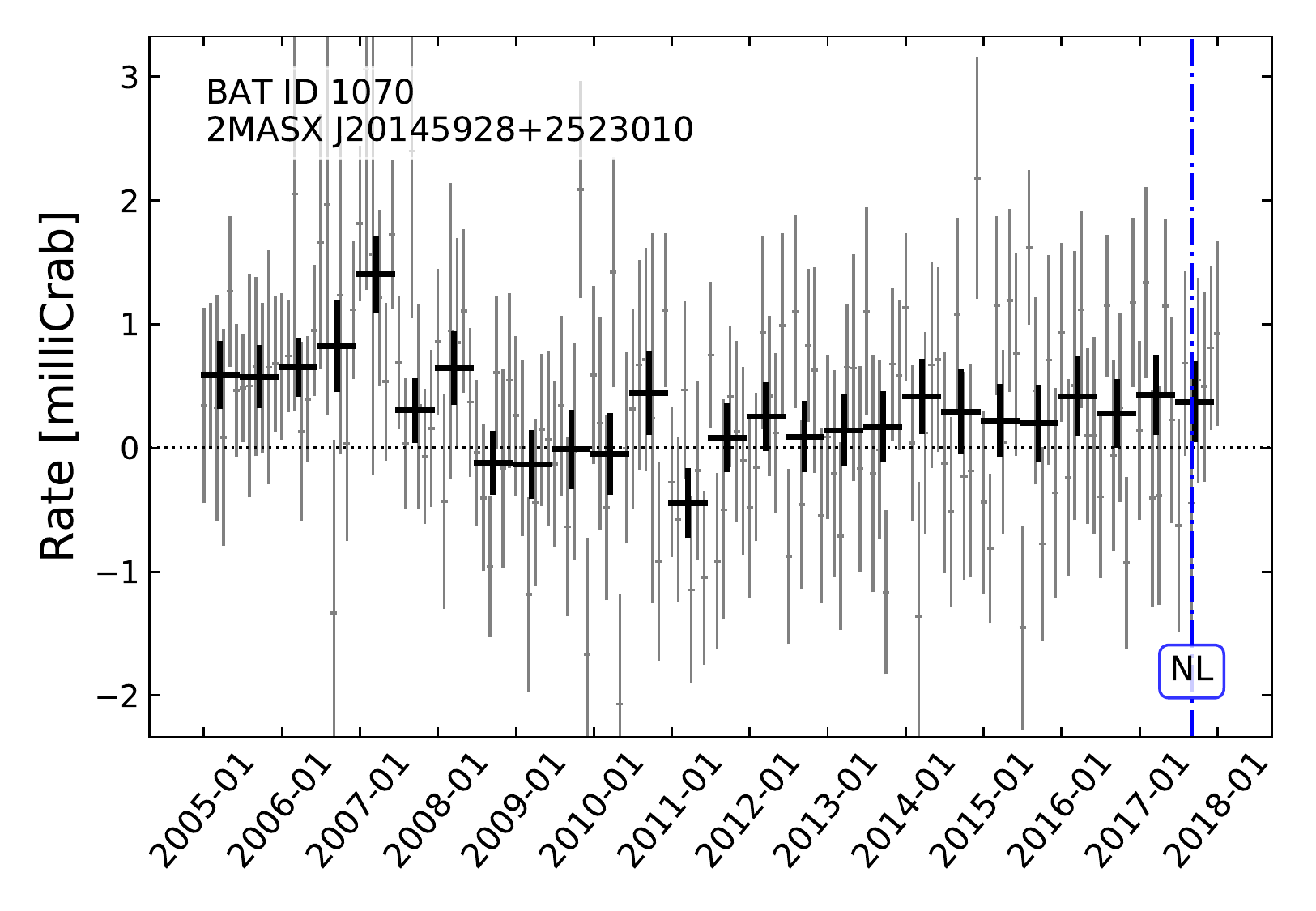}
    \caption{\textit{Swift}-BAT 14--195\,keV light-curves spanning December 2004 to December 2017 inclusive for the eight CL AGN presented in Section~\ref{sec:new_CSAGN}. Individual months are shown in gray, and rebinned to 6-month intervals  in black.
    Epochs labelled in red as `BL' correspond to the dates of optical spectra in which broad Balmer emission is seen; epochs in blue labelled as `NL' are those where only narrow Balmer lines are seen.
    }
    \label{fig:BAT_LCs}
\end{figure*}

In this section we present eight newly identified CL events and compare their optical BASS spectra (Table~\ref{tab:CS_AGN} and Fig.~\ref{fig:specs}) with their 14--195\,keV light-curves from the BAT 157-month catalogue (Fig.~\ref{fig:BAT_LCs}).

\begin{table*}
	\centering
	\caption{BASS spectra for the eight CL AGN discussed in Section~\ref{sec:new_CSAGN}.
	`B': broad line observed, `N': no broad line observed, `-': no spectral coverage. 
	`Type' refers to the estimated type classification using the criteria described by \citet{1981ApJ...249..462O}.}
	\label{tab:CS_AGN}
	\begin{tabular}{lllccccr} % four columns, alignment for each
		\hline
		BAT ID & SWIFT Name & Counterpart Name & Observation dates & \hb & \ha & Type & Instruments\\
		% & Counterpart Name
		\hline
		BAT\,72 & SWIFT\,J0123.8--3504 & NGC\,526A  
		  & 2009-07-20 & N & N & 2 & CTIO/RC\\
		&&& 2016-09-12 & N & B & 1.9 & duPont/BC \\
		&&& 2018-08-24 & N & B & 1.9 & VLT/Xshooter\\
		\hline
		BAT\,184 & SWIFT\,J0333.6--3607 & NGC\,1365 
		  & 2010-09-17 & N & N & 2 & CTIO/RC \\
		&&& 2013-12-10 & B & B & 1.5 & VLT/Xshooter\\
		&&& 2017-06-21 & B & - & - & VLT/FORS2 \\
		&&& 2021-12-12 & N & N & 2 & Magellan/MagE \\
		\hline
		BAT\,280 & SWIFT\,J0528.1--3933 & ESO\,306\,--\,IG\,001
		  & 2016-03-14 & N & N & 2 & duPont/BC \\
		&&& 2017-07-19 & N & - & - & VLT/FORS2 \\
	    &&& 2019-09-02 & B & B & 1.5 & VLT/Xshooter \\
		\hline
		BAT\,349 & SWIFT\,J0655.8+3957 & UGC\,03601
		  & 1999-02-14 & B & B & 1.5 & KPNO/Goldcam \\
		&&& 2008-12-07 & N & N & 2 & KPNO/Goldcam \\
		&&& 2018-09-10 & N & B & 1.9 & Palomar/DBSP \\
		\hline
		BAT\,757 & SWIFT\,J1508.8--0013 & Mrk\,1393
		  & 2001-03-22 & N & N & 2 & APO/SDSS \\
		&&& 2022-05-31 & B & B & 1 & LCO/FLOYDS \\
		\hline
		BAT\,981 & SWIFT\,J1830.8+0928 & CGMW\,5\,--\,04382
		  & 2010-04-05 &  N & N & 2 & Perkins/DeVeny\\
		&& (LEDA\,2808003)
		  & 2014-06-05 &  N & B & 1.8 & VLT/Xshooter \\
		&&& 2016-07-11 &  B & B & 1.5 & Palomar/DBSP \\
		\hline
		BAT\,1037 & SWIFT\,J1926.9+4140 & 2MASX\,J19263018+4133053
	  	  & 2010-05-28 &  N & N & 2 & Perkins/DeVeny \\
		&& (LEDA\,2182842)
		  & 2015-08-11 &  N & B & 1.9 & Palomar/DBSP \\
		&&& 2018-03-27 &  N & - & - & Palomar/DBSP \\
		\hline
		BAT\,1070 & SWIFT\,J2015.2+2526 & 2MASX\,J20145928+2523010
		  & 2017-08-05 &  N & N & 2 & Palomar/DBSP \\
		&&& 2017-08-27 &  N & N & 2 & Palomar/DBSP \\
		&&& 2019-06-11 &  N & B & 1.9 & Palomar/DBSP \\
		\hline
	\end{tabular}
\end{table*}

\begin{figure*}
    \centering
    \includegraphics[width=2\columnwidth, clip=on, trim={0 30 0 0}]{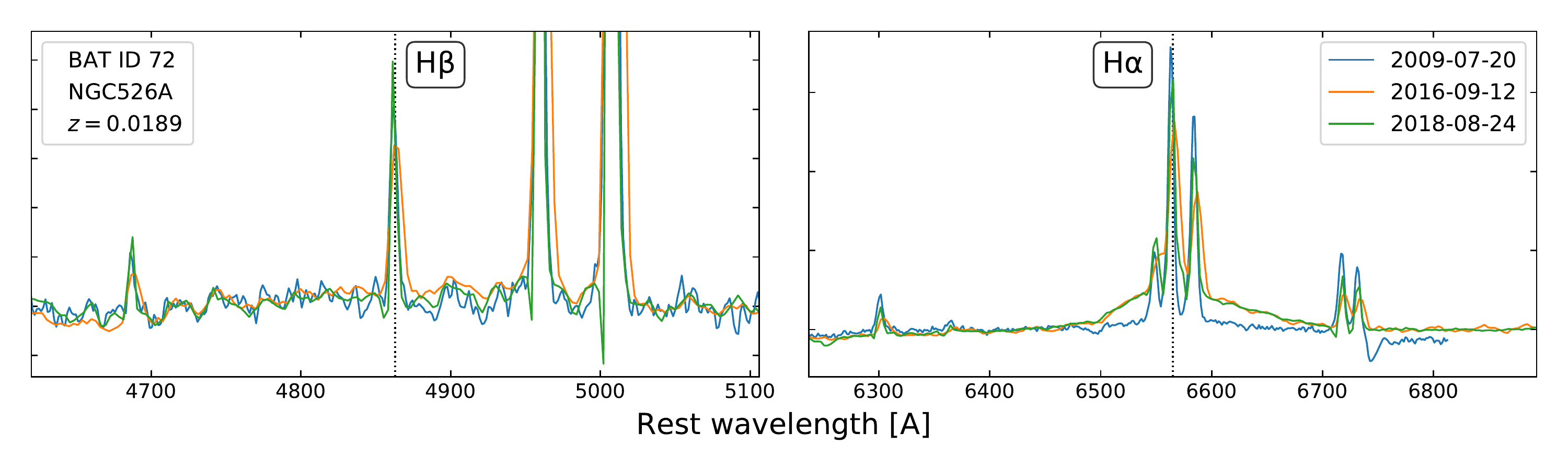}
    \includegraphics[width=2\columnwidth, clip=on, trim={0 30 0 0}]{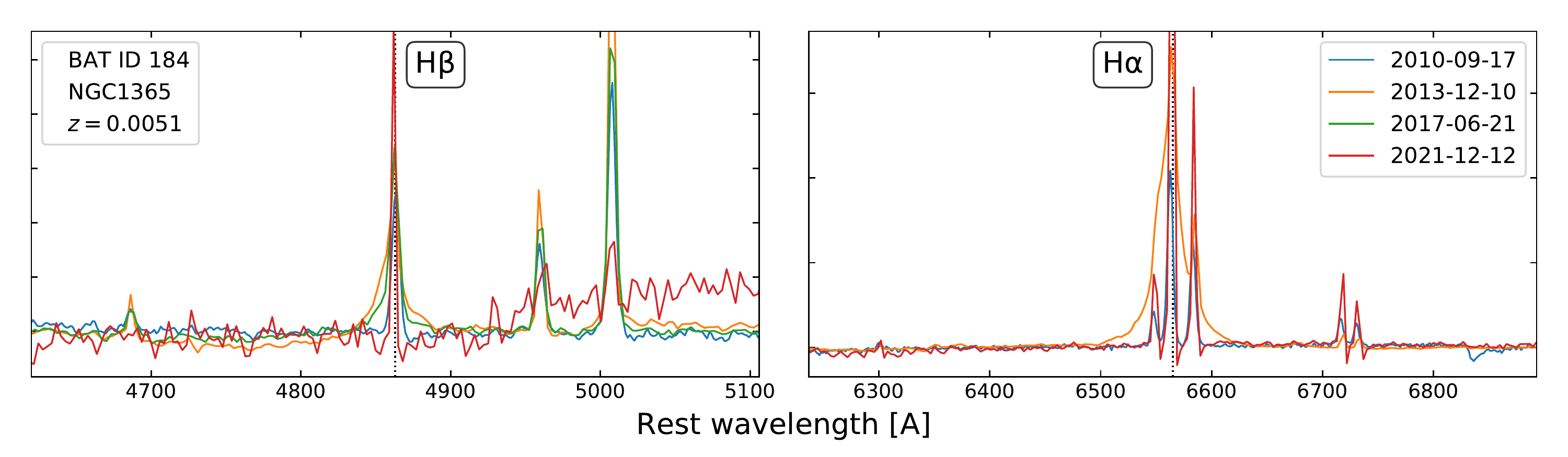}
    \includegraphics[width=2\columnwidth, clip=on, trim={0 30 0 0}]{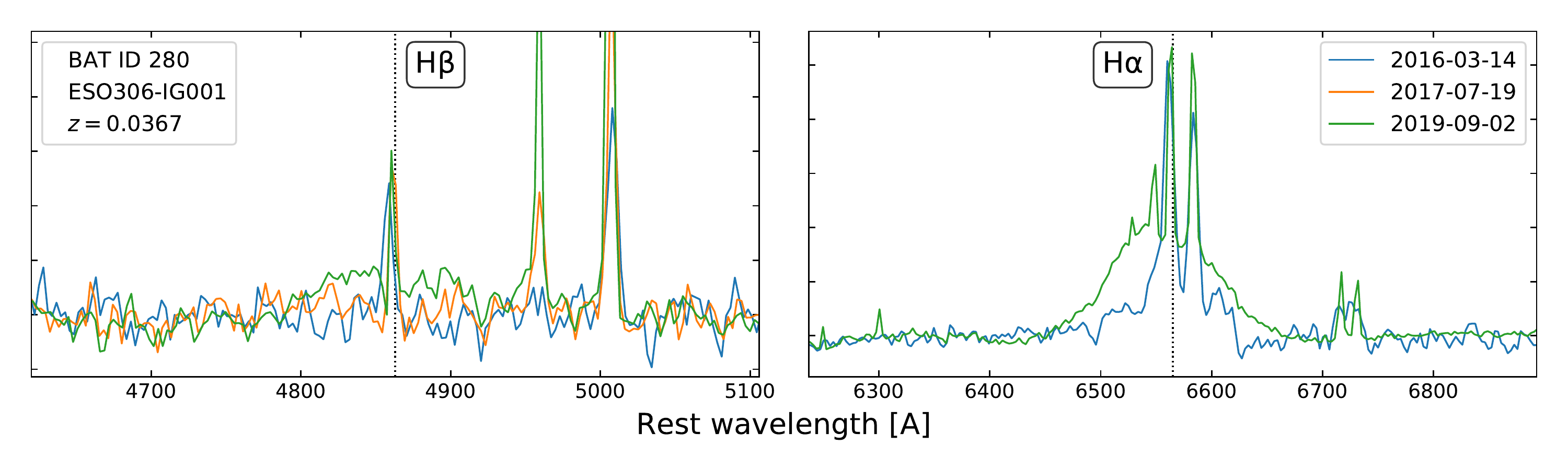}
    \includegraphics[width=2\columnwidth, clip=on, trim={0 0 0 0}]{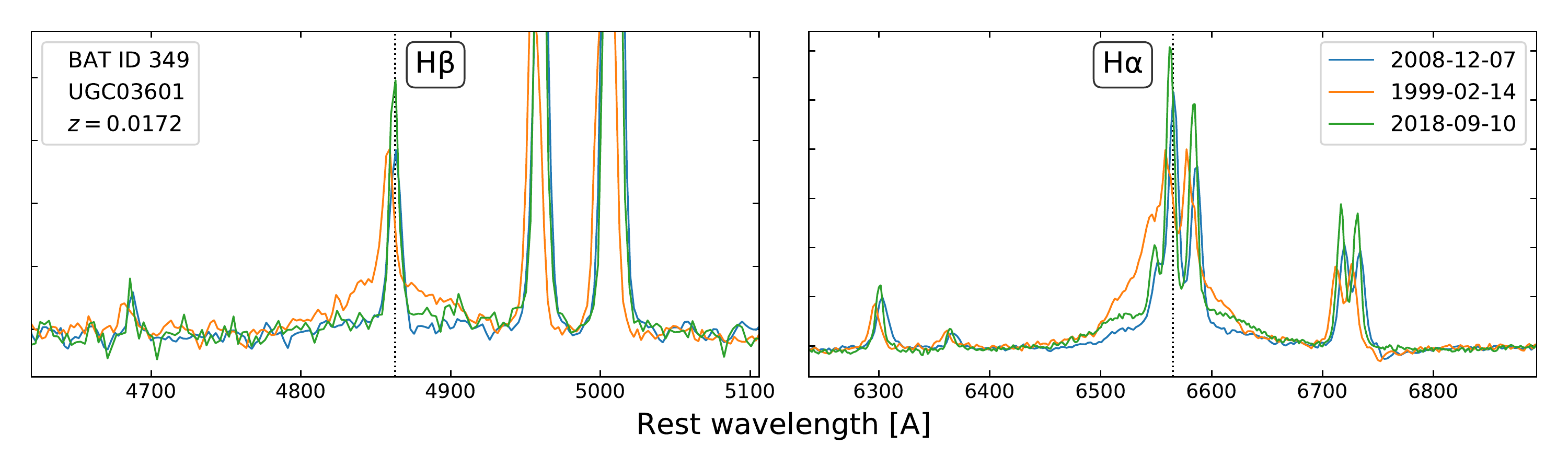}
    \caption{Flux density spectra for BASS AGN with CL events newly identified in this work.
    Counts have been normalised using the median flux density across the 4500-5200 and 6000-7000\,\AA\ spectral windows to allow a direct comparison of the emission line morphologies. Dotted lines show the rest-frame wavelengths of Balmer  \hb\ $\lambda$4861 and  \ha\ $\lambda$6563. }
    \label{fig:specs}
\end{figure*}
\begin{figure*}
    \centering
    \includegraphics[width=2\columnwidth, clip=on, trim={0 30 0 0}]{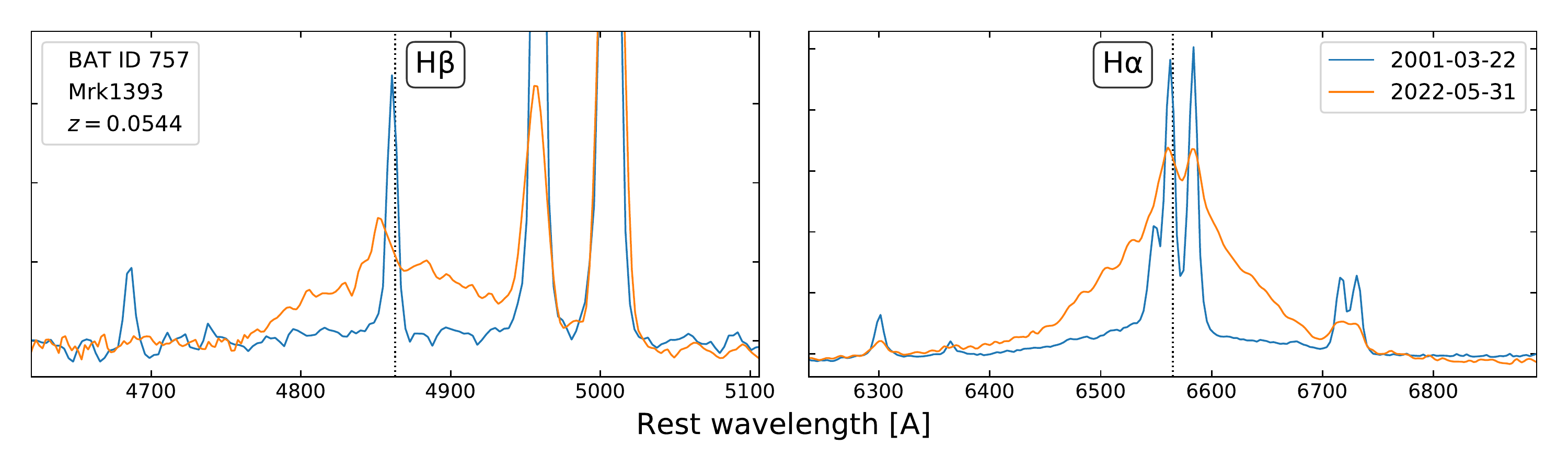}
    \includegraphics[width=2\columnwidth, clip=on, trim={0 30 0 0}]{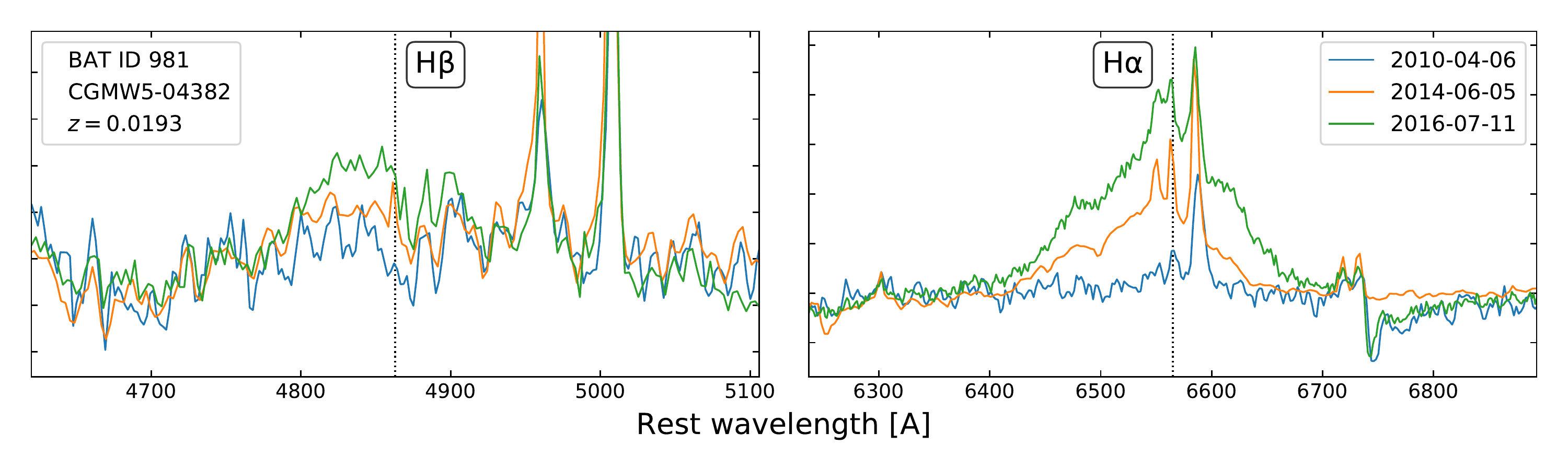}
    \includegraphics[width=2\columnwidth, clip=on, trim={0 30 0 0}]{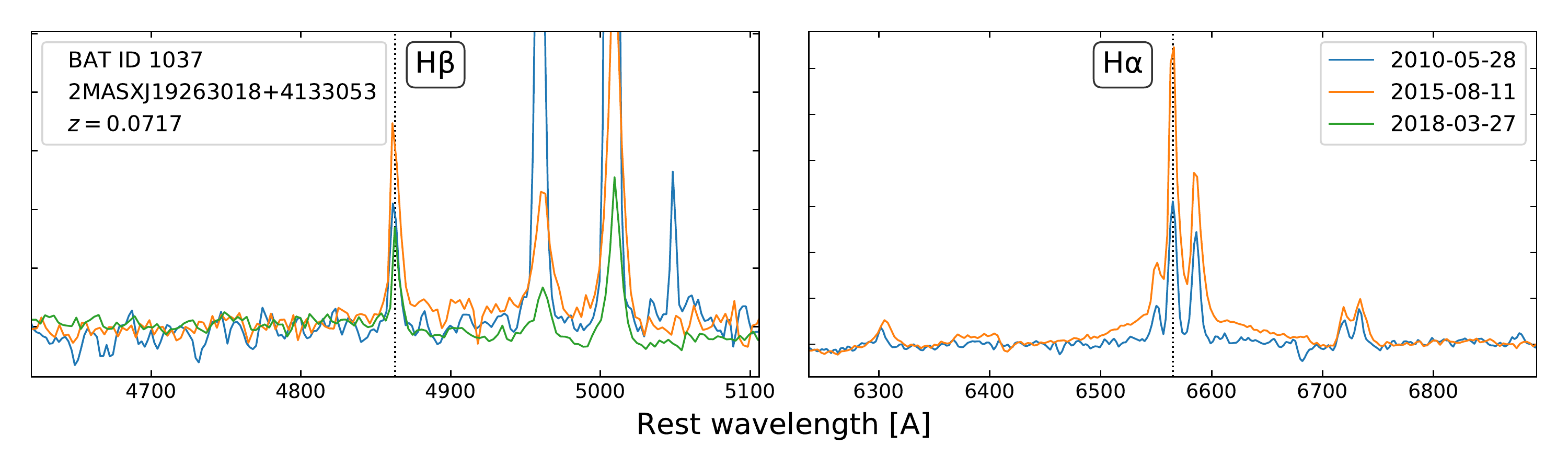}
    \includegraphics[width=2\columnwidth, clip=on, trim={0 0 0 0}]{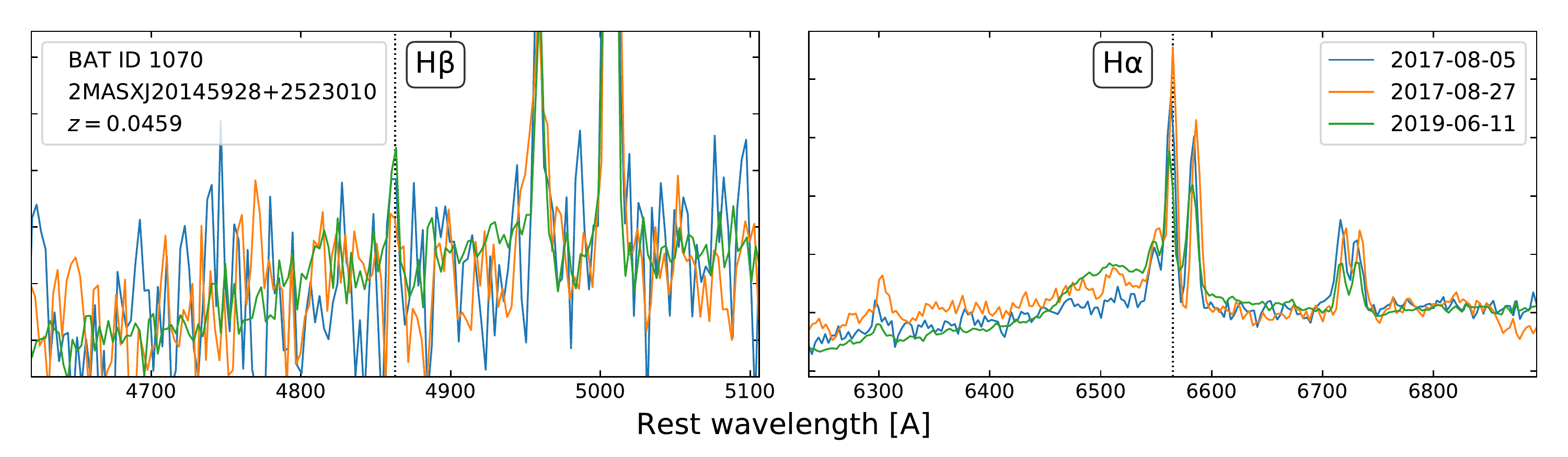}
    \contcaption{}
\end{figure*}
\begin{figure*}
    \centering
    \includegraphics[width=\columnwidth, clip=on, trim={0 15 0 0}]{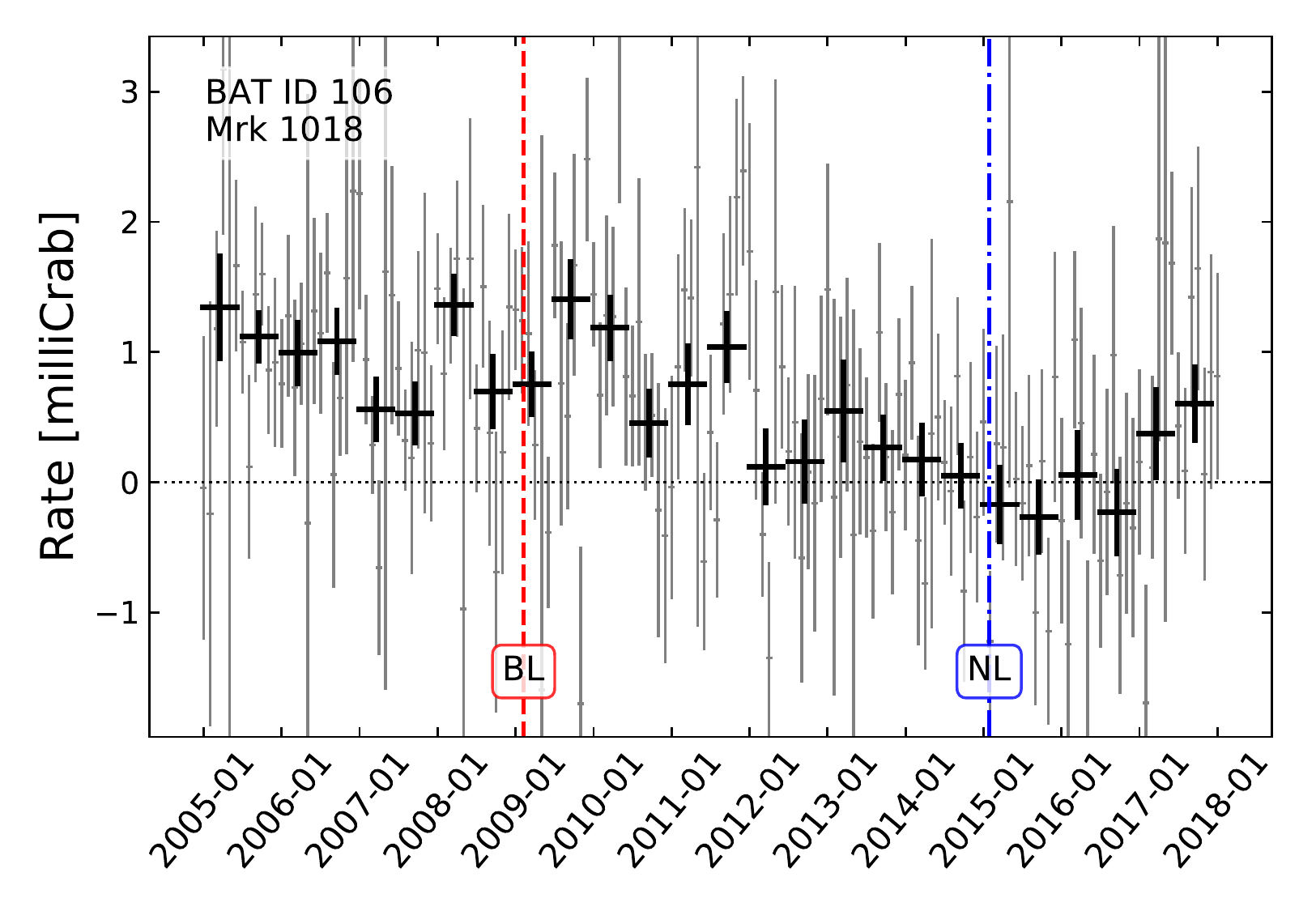}
    \includegraphics[width=\columnwidth, clip=on, trim={0 15 0 0}]{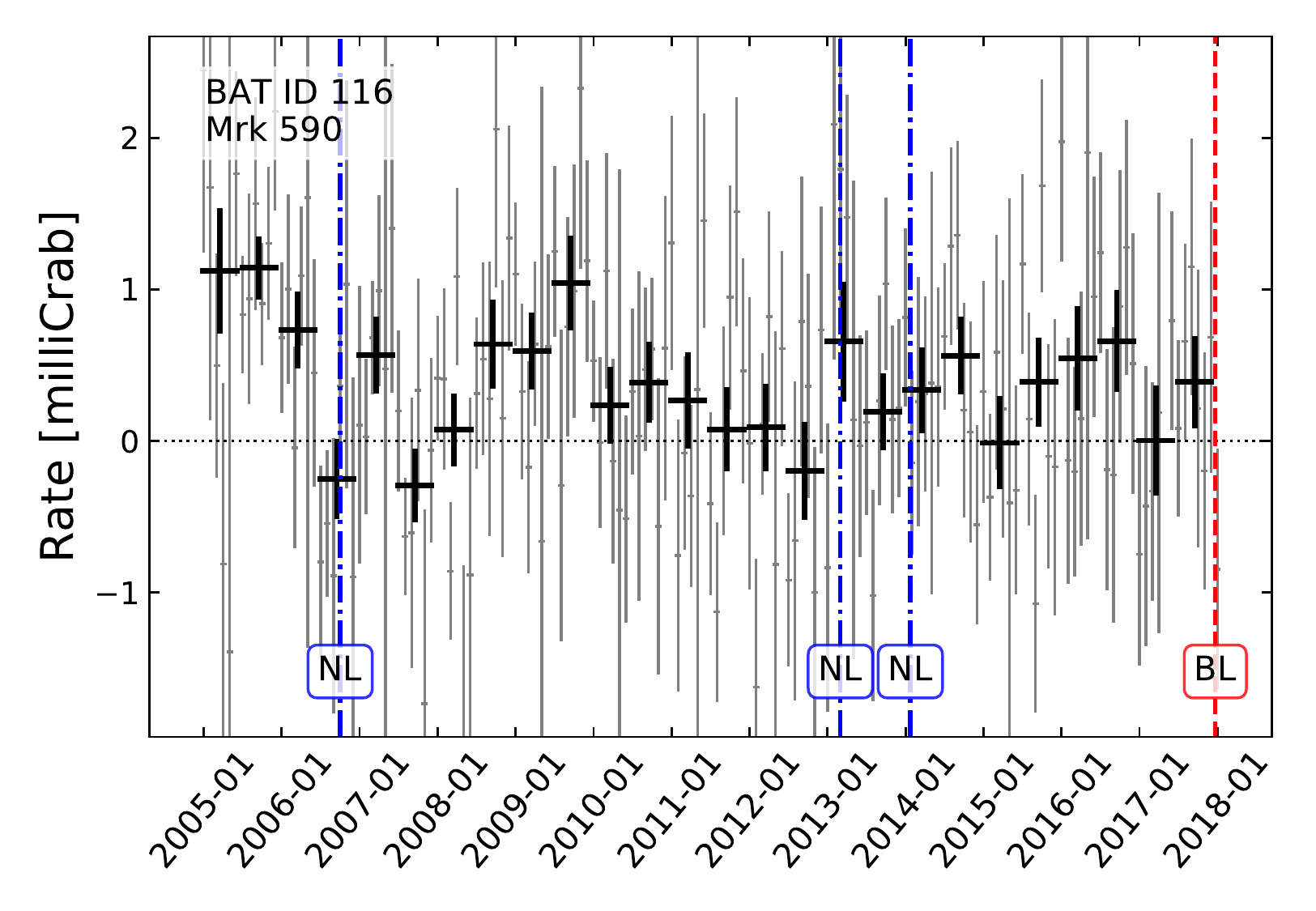}
    \includegraphics[width=\columnwidth, clip=on, trim={0 15 0 0}]{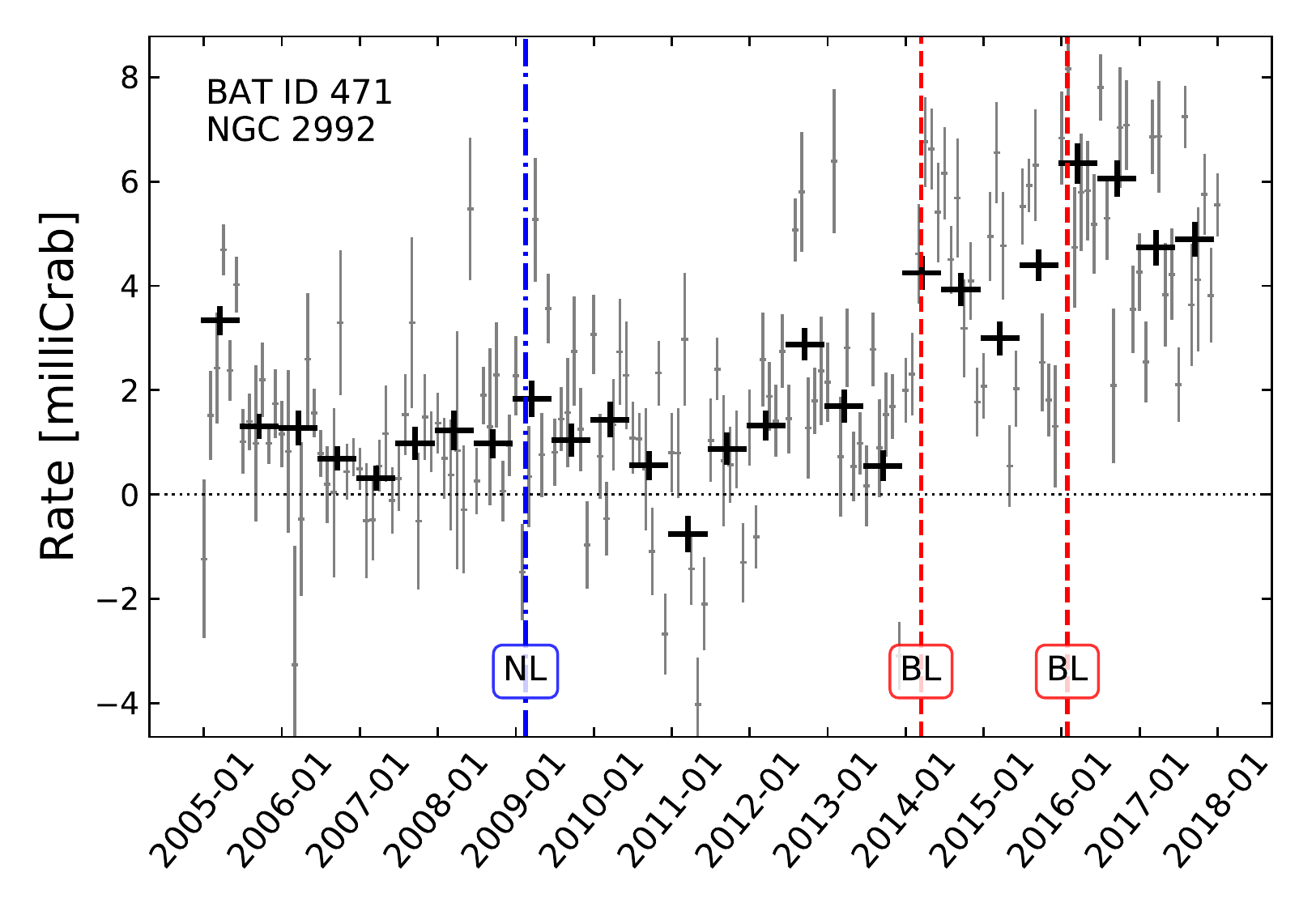}
    \includegraphics[width=\columnwidth, clip=on, trim={0 15 0 0}]{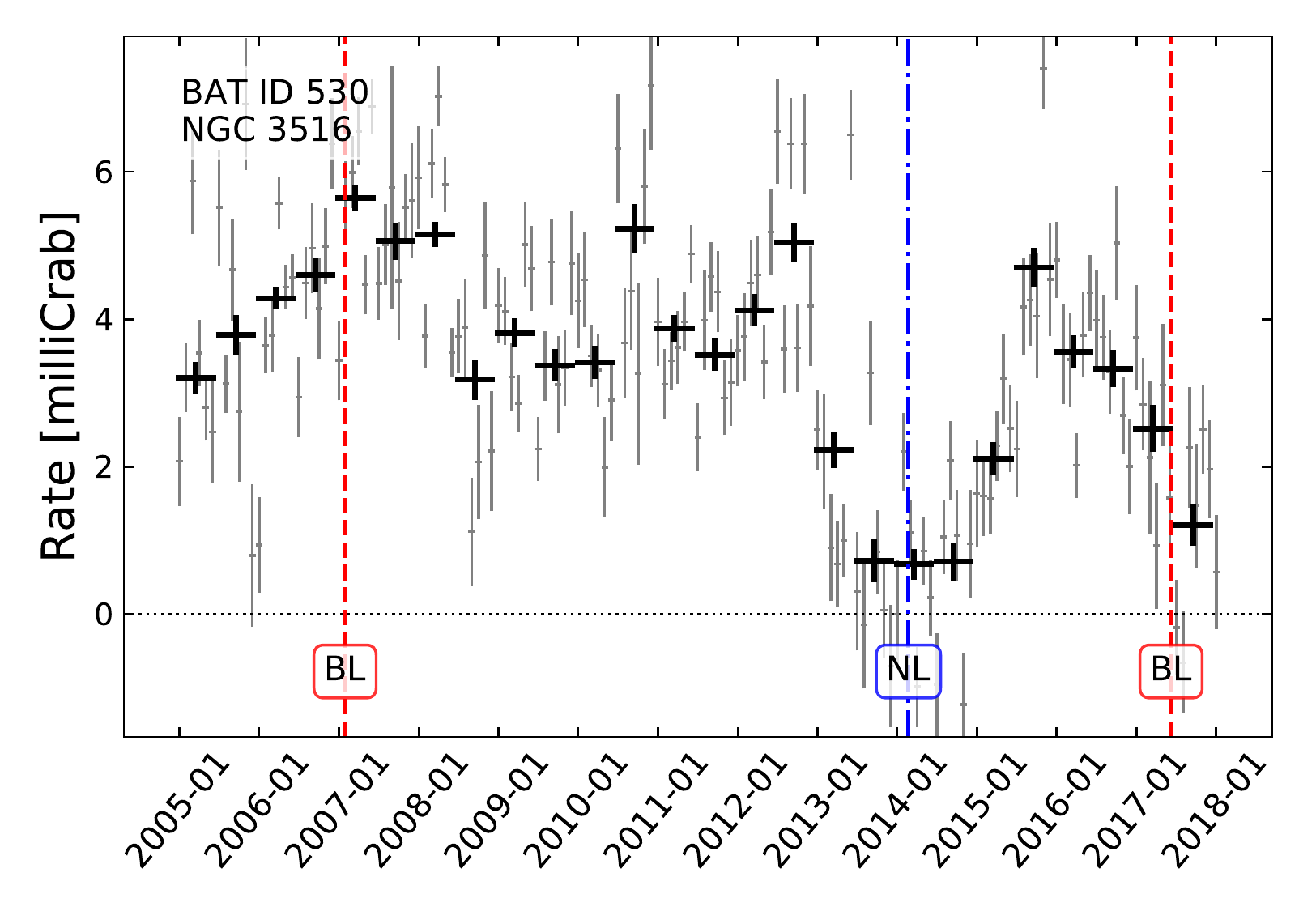}
    \includegraphics[width=\columnwidth, clip=on, trim={0 15 0 0}]{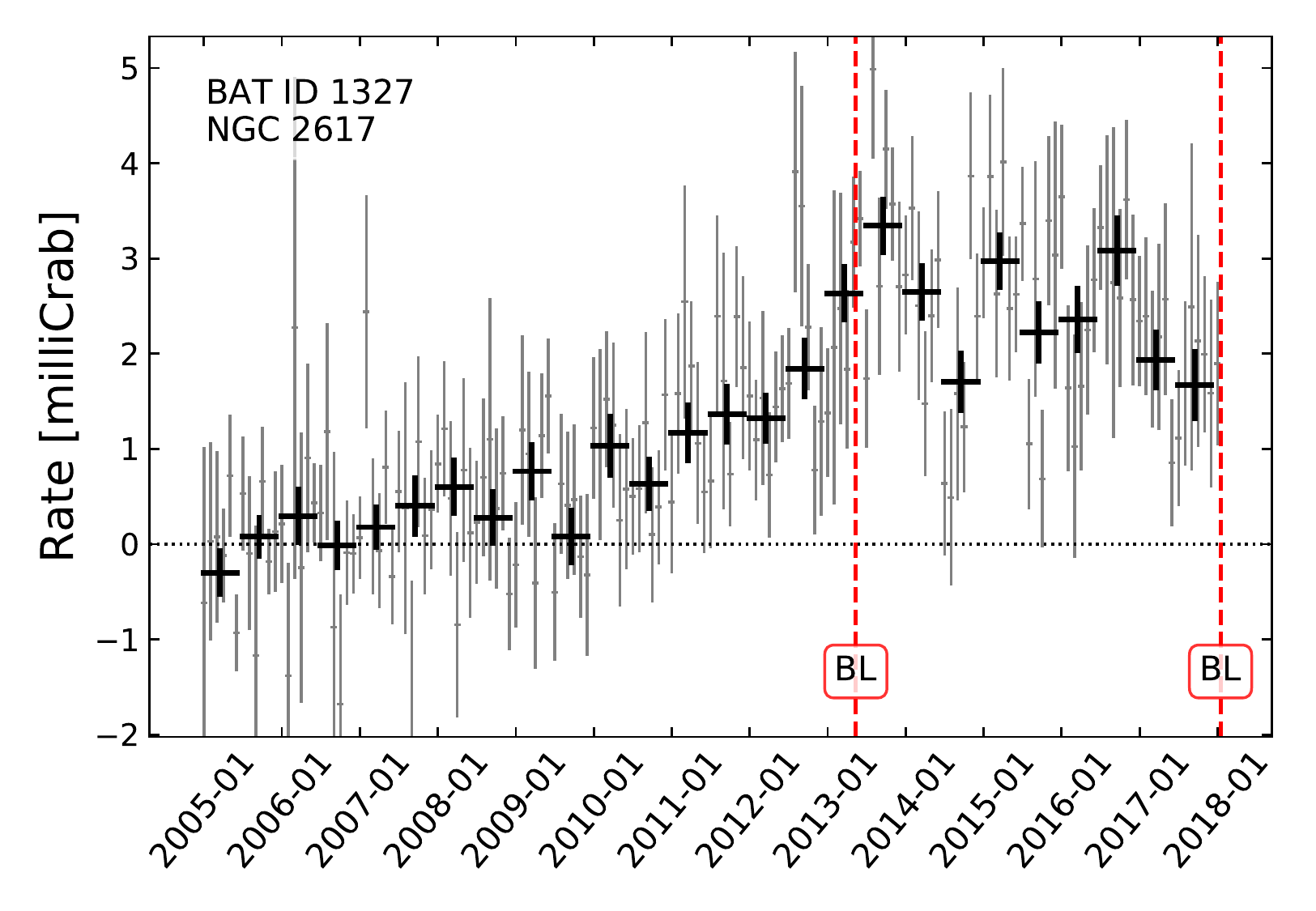}
     \caption{\textit{Swift}-BAT 14--195\,keV light-curves spanning December 2004 to December 2017 inclusive for the five known CL AGN discussed in Section~\ref{sec:known_CSAGN}. Individual months are shown in gray, and rebinned to 6-month intervals  in black.
    Epochs labelled in red as `BL' correspond to the dates of optical spectra in which broad Balmer emission is seen; epochs in blue labelled as `NL' are those where only narrow Balmer lines are seen.}
    \label{fig:BAT_LCs2}
\end{figure*}

\subsubsection{NGC\,526A}

NGC\,526A (BAT\,72) was observed as a type 2 source with only narrow Balmer emission lines in 2009 July. In early 2010, an increase was seen in the BAT flux by around a factor of two. In 2016 and 2018, the source is observed with a broad component in the \ha\ emission line and so we consider it to be a `turn-on' CS AGN, although we note that no corresponding change is detected in \hb.

\subsubsection{NGC\,1365}

We present a newly discovered `turn-off' event in the CL AGN NGC\,1365 (BAT\,184).
NGC\,1365 was observed to show broad emission lines in both \ha\ and \hb\ in 1993 August \citep{1999A&A...346..764S}, but in a spectrum taken in 2009 January
\citet{2010ApJ...725.1749T} report only narrow \hb, with any broad \ha\ emission being very weak, suggesting that the source `turned off' between 1993 and 2009.
In BASS DR1 we released a 2010 September spectrum which showed no change from 2009 January, with NGC\,1365 still displaying only narrow \ha\ and \hb.
However, we then see broad Balmer lines in our 2013 December and 2017 June spectra, suggesting that the source has turned back on between 2010 and 2013.
\citet{2017MNRAS.464.1783O} find broad near-infrared Paschen lines in 2011 October observations with ISAAC, suggesting that either the turn-on event happened rapidly between 2011 September and 2011 October, or that the type 2 behaviour observed in the optical in 2009-10 was due to dust obscuration which attenuated the broad Balmer lines \citep[e.g.][]{1995ApJ...440..141G}.
We note that \citet{2016MNRAS.459.4485L} found broad \ha\ in 2013 January IFU observations with GMOS, as did \citet{2018A&A...619A..74V} in 2014 October observations with MUSE, who also found a kilo-parsec--scale bi-conical outflow \citep[see also][]{2022MNRAS.511.2105K}.
We re-observed this source with Magellan MagE in 2021 December, identifying only narrow \ha\ and \hb, suggesting that the source once again turned off between 2014 October and 2021 December in a newly discovered CL event.
 
 NGC\,1365 is well-known to have previously displayed rapid changes in its X-ray spectral properties, from Compton-thick to Compton-thin and back again within just six weeks in 2002 and 2003. The rapid nature of this change led to its attribution to variable obscuration
\citep{2003MNRAS.342..422M, 2005ApJ...623L..93R, 2013MNRAS.436.1615M}.
Further X-ray follow-up found NGC\,1365 in a high-flux state with low column absorption in 2013 January \citep{2014ApJ...795...87B}, although recent work has suggested that the X-ray observations can also be explained by a variable accretion rate leading to changes in the coronal geometry \citep{2022A&A...662A..77M}.
This variable obscuration may lead to additional uncertainties in the intrinsic luminosity traced by the BAT light-curve.
Between 2004 and 2013, the flux in the BAT light-curve flickers by around a factor of three, although no obvious correlation can be drawn between the changes in the optical classifications and the peaks and troughs in the X-ray light-curve.

\subsubsection{ESO\,306\,--\,IG\,001}

ESO\,306\,--\,IG\,001 (BAT\,280) is an interacting galaxy pair, separated by around 20\,arcsec. We have verified that all the spectra used in our analysis are of the southern, active galaxy nucleus, which is clearly identifiable by virtue of its bluer colours in optical imaging and redder colours in WISE. 
In 2019 September, we identified the appearance of broad \ha\ emission which was not present in 2016 March, and moreover the appearance of broad \hb\ emission which was not present in 2017 July.
The BAT light-curve shows tentative evidence for an  increase in flux in the second half of 2017, during the last 6 months of currently available data. Future data releases from \textit{Swift}-BAT should provide X-ray fluxes covering the period from 2017 to 2019 when the change was observed in the optical spectra.

\subsubsection{UGC\,03601}

UGC\,03601 (BAT\,349) shows emission from broad \ha\ and \hb\ lines in 1999. In 2008 and 2018, emission from broad \ha\ is significantly weaker, and no broad component is visible in \hb. The BAT light-curve shows the X-ray flux dropping by a factor of $\approx$3 from 2006 to 2008, and staying at a lower level through to the end of the 157-month survey in 2018.

\subsubsection{Mrk\,1393}

Mrk\,1393 (BAT\,757) was observed to show galaxy-dominated continuum in 1993, 2001, 2005 \citep{2009AJ....137.4002W}, with no broad \hb\ present in the 2001 March spectrum from SDSS, leading to it being classified as a type 1.9 in BASS DR1.
However, a 1984 spectrum presented by \citet{1988MNRAS.230..639M} shows broad Balmer emission, suggesting the source turned off between 1984 and 1993.
 X-ray observations of Mrk\,1393 were presented by \citet{2009AJ....137.4002W}, who discussed a scenario in which variable obscuration in the optical and X-rays could be due to dusty material disrupted from the circumnuclear torus. \citet{2009AJ....137.4002W} also detected weak broad \hb\ in their 2005 September spectrum, with a line-of-sight extinction of 0.6\,mag estimated from the Balmer decrement.
A new, previously unpublished spectrum taken in 2022 May with LCO/FLOYDS shows a much brighter and bluer continuum, along with much stronger  broad \ha\ and \hb\ emission lines consistent with an unobscured line-of-sight to the BLR.
We therefore classify this as a new `turn-on' event which has taken place between 2005 and 2022. 
Recent photometry from the Zwicky Transient Facility (ZTF; \citealt{2019PASP..131a8002B, 2019PASP..131a8003M}) shows a brightening of almost two magnitudes in 2021 in the \textit{g} and \textit{r} bands (ZTF18acxcttu; AT\,2019aahm; \citealt{2020TNSTR3602....1F}), consistent with the appearance of an AGN-dominated continuum.

\subsubsection{CGMW\,5-04382}

CGMW\,5-04382 (LEDA\,2808003; BAT\,981) shows a brightening of broad line emission in both \ha\ and \hb\ between 2010 April (when it  displayed only narrow emission lines) and 2014 June, and again between 2014 June and 2016 July. We therefore classify it as a turn-on CL AGN.
X-ray spectral analysis by \citet{2017ApJS..233...17R} found a column density of $N_\textrm{H}=10^{23.2}$, suggesting this source was X-ray obscured when the source was observed by \textit{Swift}-XRT in 2007 September. 
The BAT light-curve is noisy with low flux rates, which is consistent with the possibility that the CL transition observed in this source is due to variable obscuration.

\subsubsection{2MASX\,J19263018+4133053}

2MASX\,J19263018+4133053 (LEDA\,2182842; BAT\,1037)
 resembled a type 2 AGN in 2010 May,  with only narrow \ha\ and \hb\ emission lines present. In 2015 August, a broad base is clearly visible in the \ha\ line, though a corresponding feature is not detected in \hb\ in either 2015 August or 2018 March.
2MASX\,J19263018+4133053 was detected in 2020 and 2021 as a variable source in optical photometry (ZTF20aazwurf; AT\,2020kcu; \citealt{2020TNSTR1365....1F}).
The BAT light-curve is also noisy with low signal-to-noise ratio even in the rebinned light-curve. 

\subsubsection{2MASX\,J20145928+2523010}

2MASX\,J20145928+2523010 (BAT\,1070) is a Compton-thick ($N_\textrm{H}=10^{24.2}$\,cm$^{-2}$) AGN which is also very red  (\textit{g}-\textit{r}\,$\approx$\,2\,mag) in optical imaging. Correspondingly, the signal-to-noise in the \hb\ region of the optical spectra is poor. In 2017 August only narrow \ha\ emission is detected, but in 2019 June a broad emission feature appears around $\lambda_\textrm{rest}=6510$\,\AA, which we interpret as broad, albeit blueshifted, \ha\ emission.
BAT\,1070 was detected as a transient source in ZTF photometry (ZTF19aawrrqy; AT\,2019aagk; \citealt{2020TNSTR3485....1D}), with a 0.7\,mag brightening in the \textit{r} band within 6 months in the second half of 2020 which has since faded.

The BAT light-curve showed a peak in the first half of 2007, which subsequently decreased to a low flux state, as observed between 2009 through the end of the 157-month BAT observations in December 2017. Future data releases from \textit{Swift}-BAT should include X-ray data covering the 2017-2019 period in which the optical transition was observed.

\subsection{Previously known CL events}
\label{sec:known_CSAGN}

The BASS catalogue contains some of the most well-known and well-studied AGN in the local universe, including many that had previously been observed to change type. 
Fairall\,9 (BAT\,73), NGC\,1566 (BAT\,216), HE\,1136-2304 (BAT\,557), KUG\,1141+371 (BAT\,565),  NGC\,4151 (BAT\,595), 3C\,390.3 (BAT\,994), NGC\,7582 (BAT\,1188), NGC\,7603 (BAT\,1189) and IRAS\,23226-3843 (BAT\,1194) have all been observed to show CL behaviour
\citep{1976ApJ...210L.117T, 1984MNRAS.211P..33P, 1985A&A...146L..11K, 1985ApJ...295L..33W,  1997A&A...324..904M, 1999ApJ...519L.123A,  2000A&A...361..901K, 2008A&A...486...99S, 2016MNRAS.461.1927P, 2019MNRAS.483..558O, 2020A&A...638A..91K, 2021MNRAS.501..916J, 2022ApJ...930...46L, 2022AN....34310080O}.
These nine AGN are shown with green circles in Fig.~\ref{fig:BHMs}.

In this section, we list a further five AGN where a change in state has been identified in the literature, and where these transitions have occurred during the course of the 157-month BAT survey (2004 December to 2017 December). We present 14--195\,keV light-curves for these objects in Fig.~\ref{fig:BAT_LCs2}.

\subsubsection{Mrk\,1018}

Mrk\,1018 (BAT\,106) was observed to turn on between 1979 and 1984 \citep{1986ApJ...311..135C}, and then turn off again between 2009 and 2015 \citep{2016A&A...593L...8M, 2016A&A...593L...9H}.
In the X-rays, there is a drop in flux from 2009 to 2015 which we associate with the turn-off CS event seen in the optical.
Mrk\,1018 was observed by SDSS in 2000 September with a type 1 spectrum which
was subsequently included in BASS DR1 and DR2.

\subsubsection{Mrk\,590}

Mrk\,590 (BAT\,116) is a well-known variable AGN and has been extensively studied in multiple wavebands. Between 1996 and 2006, the source transitioned from type 1 to type 2, as reported by \citet{2014ApJ...796..134D}, who also showed that broad Balmer emission was not present in 2013 February, 2013 December or 2014 January. \citet{2018ApJ...866..123M} reported a subsequent brightening in the ultraviolet continuum and the presence of broad \ion{Mg}{II}\,$\lambda2800$ emission in \textit{Hubble Space Telescope} observations obtained in 2014 November.
A BASS DR2 observation with Xshooter in 2017 December clearly show the re-appearance of broad \ha\ and broad \hb\ emission, in agreement with the MUSE observations of \citet{2019MNRAS.486..123R} who found broad Balmer emission in 2017 October and November.
The BAT light-curve is noisy and there is no clear transition in the ultra-hard X-ray flux, although recent analysis of  X-ray spectra from \textit{NuSTAR} and \textit{Swift}-XRT  \citep{2022ApJ...937...31G} shows an increase in the Eddington ratio from $7.5\times10^{-3}$ 
in 2016 February to $1.9\times10^{-2}$ in  2018 October, around the time the source turned on in the optical spectra.

\subsubsection{NGC\,2992}

NGC\,2992 (BAT\,471) is an interacting galaxy with a bi-conical ionized gas outflow \citep{2001AJ....121..198V, 2022MNRAS.511.2105K} and a rich archive of observations analysed in detail by \citet{2021MNRAS.508..144G}.
In brief, NGC\,2992 was observed to lose its broad \ha\ emission in 1994, and then to have returned by 1999 \citep{1999ApJ...511..686A, 2000A&A...355..485G}. 
These variations were shown to correlate with the X-ray flux: 
from the high state observed by \citet{1982ApJ...256...92M}, the X-ray flux declined to a minimum in 1994, before returning to the original level in 1999 \citep{2000A&A...355..485G}.
The X-rays were further shown to undergo a changing-obscuration event by \citet{2003MNRAS.342..422M}.

A BASS spectrum taken in 2009 January shows no broad \hb\ or \ha\ emission, suggesting that the source had returned to type 2. However, in 2014 February and 2016 January, there is evidence for a broad \ha\ emission component which is not seen in 2009 \citep[see also][]{2020A&A...634A.114C}. The BAT light-curve shows an increase by a factor of $\approx$6 from 2009 through to 2016, which we associate with a further `turn-on' CL event.
Near-infrared observations in 2012 January show weak broad (and strong narrow) emission in Paschen\,$\beta$ and \ion{He}{I}, which might suggest the continuing presence of a BLR that is obscured by dust at optical wavelengths \citep{2017MNRAS.464.1783O}.
However, \citet{2021MNRAS.508..144G} showed that the intrinsic (absorption-corrected) X-ray flux varies by a factor of 40 between 2010 and 2020, and that the broad \ha\ line strength is well-correlated with these variations in the intrinsic luminosity. \citet{2021MNRAS.508..144G} are therefore able to rule out changes in obscuration as an explanation for the recent transitions in the optical spectra of NGC\,2992, instead classifying it as a CS AGN.

\subsubsection{NGC\,3516}

NGC\,3516 (BAT\,530) was observed to change between type 1 in 2007 and type 2 in 2014, with broad \ha\ re-appearing by 2017 \citep{2019MNRAS.485.4790S}.
Our BASS DR1 spectrum from 2009 April shows strong broad lines, meaning that the turn-off event happened between 2009 and 2014.
The BAT light-curve shows a peak in 2007 which drops significantly to a minimum in 2014, before recovering in 2016. We associate these changes in the X-ray flux with the respective turn-off and turn-on CS events in NGC\,3516.
Further follow-up observations have been reported by \citet{2021MNRAS.505.1029O}, showing a brightening of broad Balmer line emission through 2020, consistent with the broad lines seen in 2019 June spectrum included in BASS DR2.
We refer the interested reader to \citet{2021MNRAS.505.1029O} for a full discussion of the history of this object.

\subsubsection{NGC\,2617}
NGC\,2617 (BAT\,1327) was classified as type 1.8 in 2003, and transitioned to type 1 in 2013 \citep{2014ApJ...788...48S, 2017MNRAS.467.1496O}. Subsequent followup \citep{2017MNRAS.467.1496O, 2021MNRAS.503.3886Y} showed the continued presence of broad Balmer line emission through 2016, similar to the type 1 behaviour seen in BASS spectra in 2017 and 2020.

NGC\,2617  was not included in BASS DR2 as it was not detected in the 70-month BAT survey \citep{2013ApJS..207...19B}. Fig.~\ref{fig:BAT_LCs2} shows a strong increase in the 14--195\,keV X-ray flux from 2010-2012, and NGC\,2617 was detected in the BAT 105-month survey \citep{2018ApJS..235....4O}. Its BAT flux remained high from 2012 through to the end of the 157-month light-curve in 2017 December. We therefore associated the increase in X-ray flux in 2010--2012 with the turn-on CS event seen in the optical spectra.

\section{Discussion}
\label{sec:discuss}

\subsection{Rate of CL events in local AGN}

Given the heterogeneity of our parent sample, which takes optical spectra from multiple sources with various different selection functions, it is not possible to draw precise conclusions about the rate of CL events from the sample discussed in this paper. However, we can instead use this population to estimate upper and lower limits on the rate of CL transitions in BASS AGN. The size of our CL sample is small and so the quoted uncertainties in the following discussion are derived assuming Poisson statistics.

From the 749 unique non-beamed $z<0.5$ AGN in the BAT 70-month catalogue,
%(i.e. excluding BAT\,1327 which is only detected in the 105-month catalogue)
we know of eight CL AGN where the change in optical type classification occurred  between November 2004 and December 2017: BAT IDs 72, 106, 116, 184, 471, 530, 981, and 1037.
This places a robust lower limit of at least $1.1\pm0.4$ per cent of local ($z\approx0.03$) AGN undergoing changes in state on observed-frame time-scales shorter than 13.1 years.

Restricting to our parent sample, we have 412 objects where we have temporal coverage such that we might have expected to observe changes in the Balmer lines. 
Taking the same eight CL objects as above gives a mid-range estimate of $1.9\pm0.7$ per cent of BASS AGN displaying CL behaviour within a 13.1 year time frame.

Including all archival results from the literature (Section~\ref{sec:known_CSAGN}), 21 BASS AGN are now known to have undergone at least one CL transition over the past 50 years.
This gives a lower limit of at least $2.8\pm0.6$ per cent (21/749) of non-beamed $z<0.5$  BASS AGN displaying CL behaviour within a 50 year period.
If all 21 of these CL AGN had been found from within the parent sample in this work, we would have derived a rate of $5.1\pm1.1$ per cent (21/412) of local AGN undergoing at least one CL event within $\approx$10-25 year time-scales. In reality this likely represents an over-estimate of the rate of CL events in BASS AGN.

These conservative constraints are consistent with the results of \citet{2022MNRAS.513L..57L}, who estimate that around 1.8 per cent of type 2 AGN transition to type 1 over the course of 15 years. However, we note that the uncertainty associated with this measurement remains large, as they extrapolate from spectroscopic follow-up observations of just six of their 30 CL candidates.
Similarly, while \citet{2016ApJ...821...33R} reported that 38 per cent of $0.02<z<0.1$ Seyfert galaxies show changes in their \hb\ flux  on 3--9 year time-scales, only $2.9\pm1.7$ per cent of their sample show the complete disappearance of broad Balmer emission, which is consistent with our results (derived through a similar approach).
\citet{2021MNRAS.tmp.3368H} also report a CL rate of $\approx$3 per cent per 15 years, which is consistent with our results.
On the other hand, \citet{2022ApJ...933..180G} found only 19 CL events from a sample of 64\,039 SDSS quasars, giving a rate of $\approx$0.03 per cent across a $\approx$10 year time-scale, some two orders of magnitude lower than the CL rate which we find in BASS AGN. However, \citet{2022ApJ...933..180G} do not claim to be complete, and their initial sample of quasars have much higher Eddington ratios compared to the BASS sample. Such objects are known to be less likely to display CL behaviour \citep{2019ApJ...874....8M}, so it is not surprising that \citet{2022ApJ...933..180G} find a lower incidence of CL behaviour.

Our inferred rate of $\approx$0.7--6.2 per cent per 10--25 years is significantly higher than those of tidal disruption events, which are predicted to occur of order  10$^{-4}$ times per year per galaxy, and are observed even less frequently \citep{2016MNRAS.455..859S, 2020SSRv..216...35S, 2021ARA&A..59...21G}. Tidal disruption of stars is therefore unlikely to be the main mechanism for our observed CL behaviour.

\subsection{Physical drivers of CL behaviour}
\label{sec:drivers}

\begin{figure}
    \centering
    \includegraphics[width=\columnwidth]{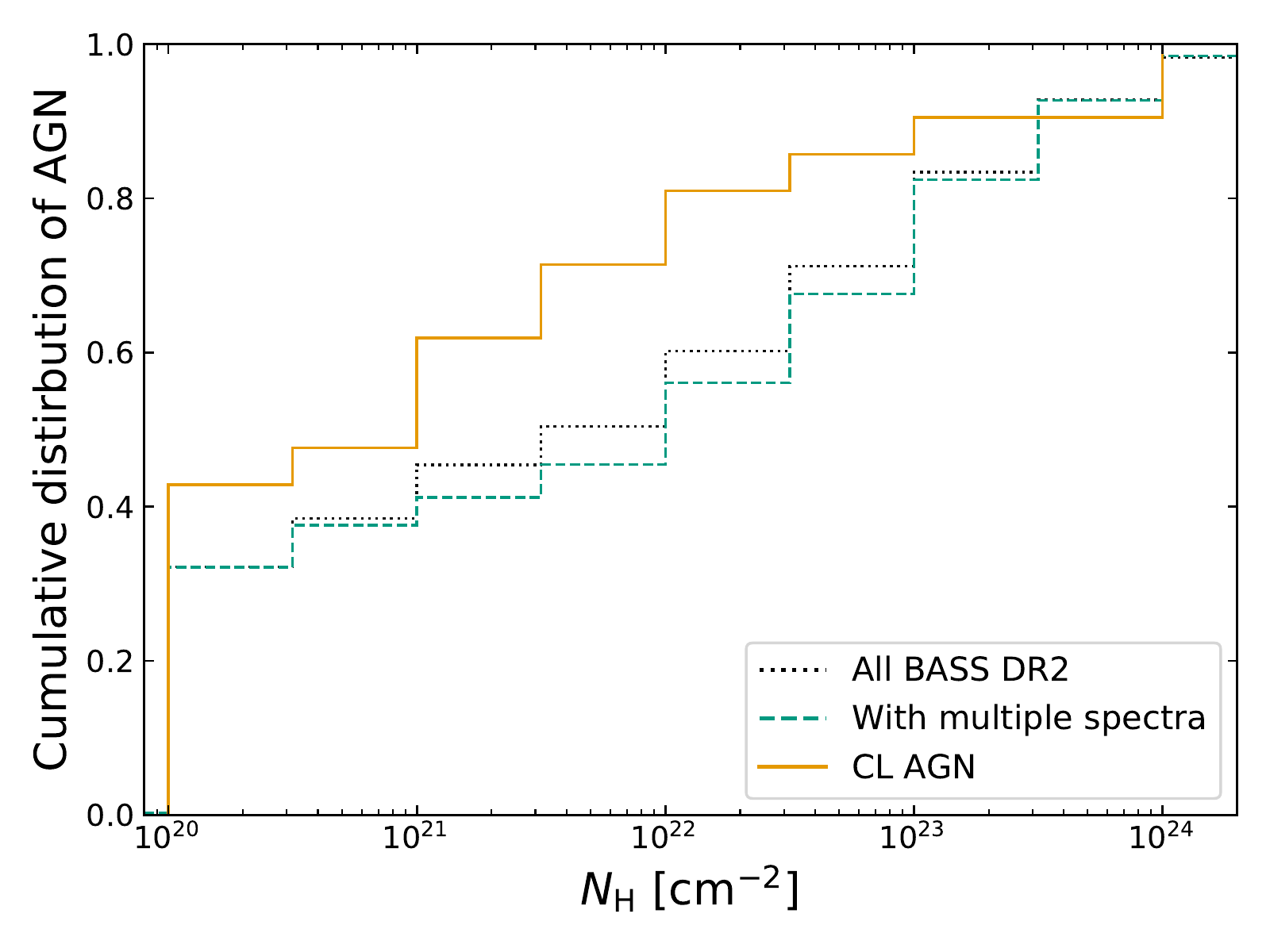}
    \caption{Cumulative distribution of the column densities along the lines of sight to BASS AGN, as measured by \citet{2017ApJS..233...17R}. 
    The CL AGN discussed in this work (solid orange line) tend to prefer lower values of $N_\textrm{H} < 10^{22}$\,cm$^{-2}$, meaning that they are less likely to be X-ray obscured than a random BASS AGN.}
    \label{fig:nh}
\end{figure}

We have identified nine %10
changes of AGN type which are constrained to have taken place during the 2004 December to 2017 December period of the 157-month BAT survey. This includes two CL events in BAT\,530 and one event each in the seven other CL AGN listed in the previous section.
%, and one event in BAT\,1327, which was not detected in the 70-month \textit{Swift}-BAT survey and so was not included in our rate estimates above.
Three out of these nine %ten 
events could be due to variable obscuration: BAT\,981 and BAT\,1070 have had X-ray column densities measured of $N_\textrm{H} = 10^{23.2}$ and $10^{24.4}$\,cm$^{-2}$ respectively \citep{2017ApJS..233...17R}, and BAT\,184 is known to have variable X-ray obscuration \citep{2003MNRAS.342..422M, 2005ApJ...623L..93R, 2013MNRAS.436.1615M}.
Five %Six 
of the six %seven 
other CL events show clear changes in their BAT light-curves, with broad-line epochs (as expected) corresponding to brighter X-ray levels compared with narrow-line epochs. The remaining event was that in BAT\,116, which has an inconclusive BAT light-curve but otherwise has been shown to undergo significant changes in its intrinsic luminosity \citep{2022ApJ...937...31G}.
These changes in X-ray luminosity (in six %seven 
out of nine %ten 
CL events) would not be expected if the optical changes were solely due to variable Compton-thin obscuration, as the 14--195\,keV energy band is less affected by changes in the line-of-sight column.

In general, the column densities from X-ray spectral analysis (Fig.~\ref{fig:nh}) tend to be lower in  CL AGN than in the overall population.
The median $N_\textrm{H}$ is $10^{21.3}$\,cm$^{-2}$ in the 21 BASS AGN which are now known to be CL AGN, compared to  $10^{22.3}$\,cm$^{-2}$ in our parent sample, consistent with a scenario in which CL AGN are mostly unobscured.
Previous work from the BASS collaboration \citep{2017ApJ...850...74K, Oh_DR2_NLR} has shown that $\approx$90 per cent of Seyfert type classifications in local AGN agree with their X-ray obscuration.
Our results therefore support a scenario in which many of the most extreme changes in observed AGN broad line emission are due to changes in the underlying accretion rate.

In Fig.~\ref{fig:BHMs} we show the distribution of our new CL AGN in the $M_\textrm{BH}$--$L/L_\textrm{Edd}$ plane.
The uncertainty on our estimated $L/L_\textrm{Edd}$ includes that associated with the assumption of a constant bolometric correction, which we estimate to be around 0.45\,dex following the scatter between $L_\textrm{14-195\,keV}$ and $L_\textrm{5100\,\AA}$ found in BASS DR1 \citep{2017ApJ...850...74K}.
With this caveat in mind, we can say that we do not observe a significant number of CL AGN with $L/L_\textrm{Edd} \gtrsim 0.1$. 
We conducted a Monte Carlo experiment, producing 100\,000 random realisations of our CL sample from the parent distribution of  $L/L_\textrm{Edd}$. Only 0.37 per cent of these realisations had $L/L_\textrm{Edd} < 0.1$ for every object, suggesting that it is unlikely that our observed lack of high accretion rate CL AGN is due to chance.
This is in agreement with the results of \citet{2019ApJ...874....8M}, who found that CL quasars are observed more often at lower Eddington ratios compared to the SDSS quasar population.
By analogy with the accretion state transitions in stellar-mass black hole X-ray binary systems, 
\citet{2018MNRAS.480.3898N} and \citet{2019ApJ...883...76R} have suggested that CS behaviour in AGN could arise due to changes in the structure of the inner accretion disc which occur around a critical value of $L/L_\textrm{Edd} \sim 0.02$.
Our observed distribution of $L/L_\textrm{Edd}$ is consistent with such mechanisms driving the majority of the CL events we find in BASS AGN.

\subsection{Time-scales of CL events}

\begin{figure}
    \centering
    \includegraphics[width=\columnwidth]{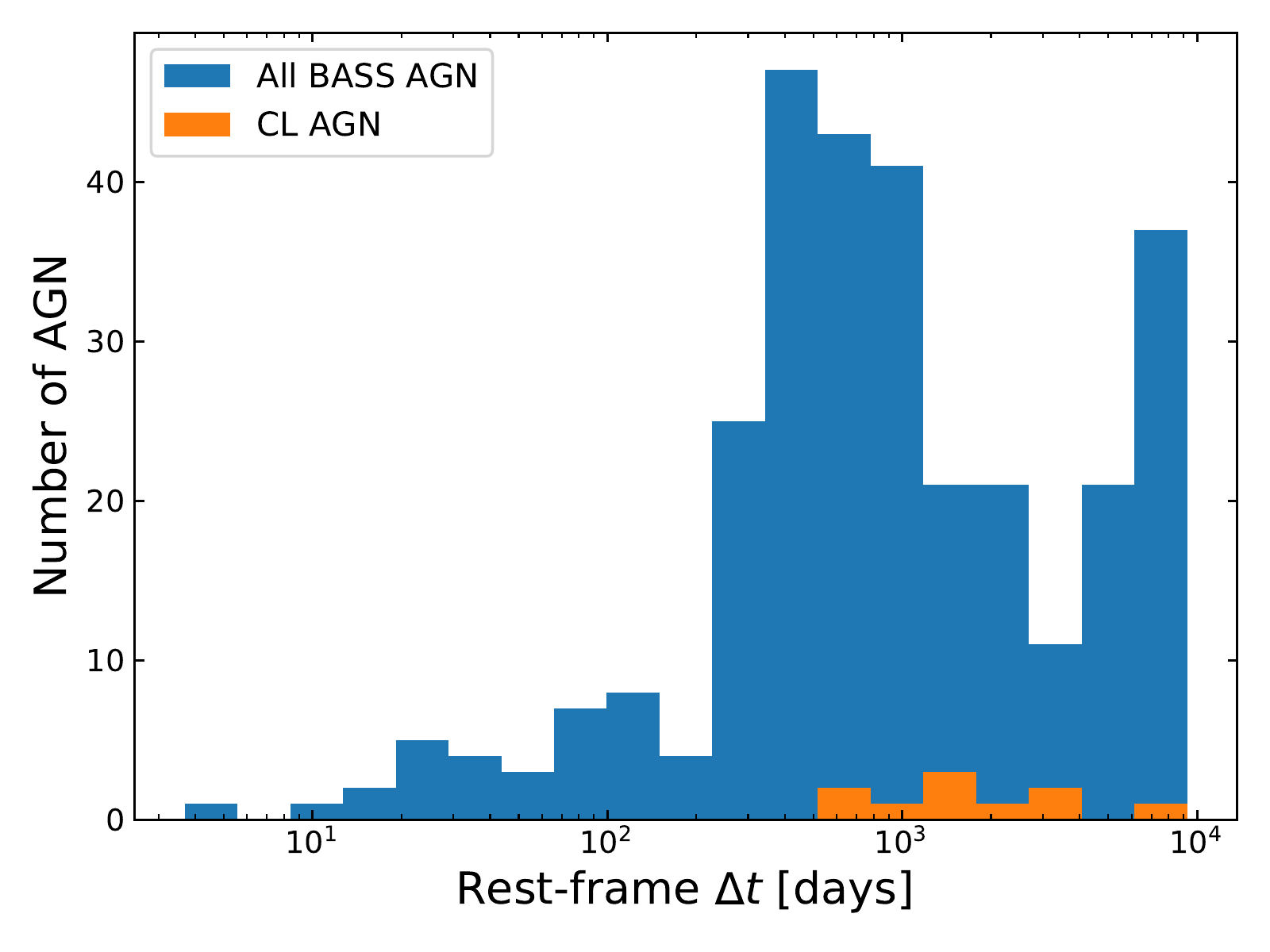}
    \caption{The distribution of $\Delta t$, the difference in time between the first and last spectral epochs, in the parent sample and the CL sample in this work. 
    Using a KS test, we are unable to reject the null hypothesis that  our CL events are drawn from the same distribution of $\Delta t$ as that our parent sample, with $p=0.069$.
    We are therefore limited by the cadence of observations in our parent sample, with most sources having at least one year between spectral epochs.}
    \label{fig:delta_t}
\end{figure}

The parent sample used in this work consisted of the non-beamed, $z<0.5$ AGN  with more than one  epoch of optical spectroscopy in BASS. 
This sample is very heterogeneous, with a variety in aperture size, spectral resolution, and  spectrophotometric calibration quality.
The cadence of the repeat observations in this sample was not chosen to support variability science, and in particular was not optimised to study CL events in any way.
Some sources have observations from more than one archival programme, while some sources were chosen for repeat observation due to their having poor signal-to-noise ratio, resolution or calibration in BASS DR1, meaning that the subset of BASS AGN which have multi-epoch spectroscopy could be biased in ways which are hard to quantify.
In Fig.~\ref{fig:delta_t} we plot the distribution of the rest-frame $\Delta t$, the difference in time between the first and last spectral epochs for each AGN in our parent sample, together with the distribution of $\Delta t$ for each pair of spectra which constrain a CL event. These $\Delta t$ serve as simple upper limits on the time-scale for the CL transitions themselves.
The parent sample is dominated by time baselines of more than one year, which limits our ability to associate time-scales shorter than $\sim$300\,days with any CL event.
Instead we see that CL transitions are identified over 3-5 year time-scales, consistent with the distribution of $\Delta t$ in our parent sample. 
However, we know that at least some CL transitions occur on much shorter timescales \citep[e.g.][]{2017ApJ...835..144G, 2019ApJ...883...94T, 2022arXiv221007258Z} which we  would miss in cases where a source transitions twice (e.g.\ off and then on again) within $\approx$300\,days.
Upcoming surveys such as SDSS-V Black Hole Mapper \citep{2017arXiv171103234K} will provide better constraints for a much larger sample of CL AGN, probing time-scales from days to years. Our work shows that continued high-cadence spectroscopic monitoring of the BASS sample would be valuable to place better constraints on the variability time-scales in the local AGN population.

\subsection{Future work}

Future work will involve analysing follow-up observations to characterise how the X-ray spectral properties have changed for the CL sources identified in this work. 
Broadband (0.3--195 keV) X-ray spectral fitting has previously been carried out for all BAT AGN \citep{2017ApJS..233...17R}, but this analysis assumed that there was no variability between the various epochs in which the broadband X-ray data was obtained.
A more sophisticated analysis will look to confirm whether the CL events we have found (especially those for which we do not have BAT light-curves) are associated with changes in the intrinsic X-ray luminosity or spectral index.

Many of the CL AGN presented herein also have light-curves available from wide-field, multi-epoch imaging surveys, both in the optical (e.g., ZTF) and the IR (\textit{WISE}; \citealt{2010AJ....140.1868W}). While these data sets generally do not span the same temporal baseline as the BAT light-curves presented above, further data collection is ongoing and will be valuable when seeking to understand more recent changes in the BAT AGN sample. Understanding the optical light-curves of the complete sample of BAT-selected AGN will further help to select and classify variable AGN in future surveys such as the Vera C. Rubin Observatory LSST \citep{2019ApJ...873..111I}.

\section{Conclusions}
\label{sec:conclusions}
%The last numbered section should briefly summarise what has been done, and describe the final conclusions which the authors draw from their work.

We have identified eight new `changing-look' events in low-redshift AGN, using spectra from the combined first and second data releases of BASS.
We have focused on the most dramatic spectral transitions, where broad Balmer lines completely appear or disappear between different epochs.
By combining with BASS AGN which were previously known to display CL behaviour, we constructed a sample of nine
CL events where the change in optical type classification has occurred during the first 157 months of \textit{Swift}-BAT operations (2004 December to 2017 December) and hence for which ultra-hard X-ray light-curves are available.
Of these nine spectral transitions, 
five display clear simultaneous changes in their ultra-hard X-ray flux from BAT.
This is consistent with a scenario where changes in the accretion disc are driving the changes seen in the optical type, as changing obscuration would need to be Compton-thick in every case to explain the extent of the changes seen in the ultra-hard X-ray emission.

The ultra-hard X-ray selection with \textit{Swift}-BAT which was used to define the BASS  sample should be less biased to the optical emission properties, leading to a parent sample which is less biased towards any particular sub-class of CL behaviour. 
However, the subset of BASS AGN with \textit{repeat} spectra which was used as the basis of this investigation includes observations from archival programmes which were selected through other criteria. These observations are heterogeneous with a range of aperture effects and different calibration processes.
Notwithstanding these limitations, we infer a CL event rate of 0.7-6.2 per cent over $\approx$15 year time-scales in local AGN, consistent with recent work in the literature.

The BASS DR2 is a comprehensive, multi-wavelength effort to understand the properties of a complete sample of low-redshift AGN, making the BAT AGN sample an ideal test-bed to better understand the physics of highly variable and changing-look AGN.

\section*{Acknowledgements}
%The Acknowledgements section is not numbered. Here you can thank helpful
%colleagues, acknowledge funding agencies, telescopes and facilities used etc. Try to keep it short.
We thank the anonymous referee for a constructive report.
MJT acknowledges support from CONICYT (Fondo ALMA 31190036) and a FONDECYT fellowship (Proyecto 3220516).
CR acknowledges support from FONDECYT (Iniciacion grant 11190831) and ANID BASAL project FB210003.
MJK acknowledges support from NASA (ADAP award NNH16CT03C).
BT acknowledges support from the European Research Council (ERC) under the European Union's Horizon 2020 research and innovation program (grant agreement 950533) and from the Israel Science Foundation (grant 1849/19).
FEB acknowledges support from the
ANID Millennium Science Initiative Program (ICN12\_009), CATA-BASAL (ACE21000 and FB210003) and FONDECYT Regular (1190818 and 1200495).
AFR acknowledges support from FONDECYT Postdoctorado Proyecto 3210157.
KO acknowledges support from the Korea Astronomy and Space Science Institute under the R\&D program (Project No. 2022-1-868-04) supervised by the Ministry of Science and ICT and from the National Research Foundation of Korea (NRF-2020R1C1C1005462).
FR acknowledges support from PRIN MIUR 2017PH3WAT (`Black hole winds and the baryon life cycle of galaxies').
RR thanks Conselho Nacional de Desenvolvimento Cient\'{i}fico e Tecnol\'ogico  (CNPq, Proj. 311223/2020-6,  304927/2017-1 and  400352/2016-8), Funda\c{c}\~ao de amparo \`{a} pesquisa do Rio Grande do Sul (FAPERGS, Proj. 16/2551-0000251-7 and 19/1750-2), Coordena\c{c}\~ao de Aperfei\c{c}oamento de Pessoal de N\'{i}vel Superior (CAPES, Proj. 0001). 

This research has made use of the NASA/IPAC Extragalactic Database (NED), which is funded by the National Aeronautics and Space Administration and operated by the California Institute of Technology.

%%%%%%%%%%%%%%%%%%%%%%%%%%%%%%%%%%%%%%%%%%%%%%%%%%
\section*{Data Availability}
%The inclusion of a Data Availability Statement is a requirement for articles published in MNRAS. Data Availability Statements provide a standardised format for readers to understand the availability of data underlying the research results described in the article. The statement may refer to original data generated in the course of the study or to third-party data analysed in the article. The statement should describe and provide means of access, where possible, by linking to the data or providing the required accession numbers for the relevant databases or DOIs.

The optical spectra presented in this article will be made available from the BASS website\footnote{\url{https://www.bass-survey.com}} as part of BASS data releases.
The \textit{Swift}-BAT light-curves presented in this article will be available from a forthcoming publication led by A.\ Lien.
%\footnote{\url{https://swift.gsfc.nasa.gov/results/bs157mon/}}


\bibliographystyle{mnras}
\bibliography{refs} % if your bibtex file is called example.bib

%%%%%%%%%%%%%%%%%%%%%%%%%%%%%%%%%%%%%%%%%%%%%%%%%%

%%%%%%%%%%%%%%%%% APPENDICES %%%%%%%%%%%%%%%%%%%%%

\appendix
%If you want to present additional material which would interrupt the flow of the main paper, it can be placed in an Appendix which appears after the list of references.

\section{Visual inspection}
\label{sec:app}

\begin{figure*}
    \centering
    \includegraphics[width=2\columnwidth, clip=on, trim={0 30 0 0}]{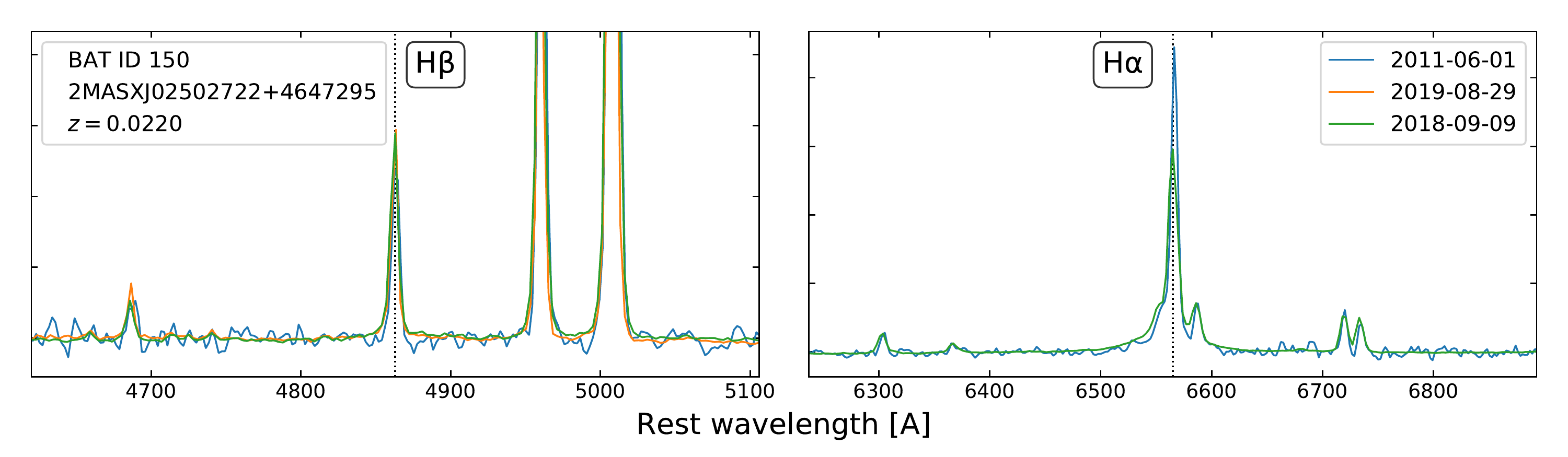}
    \includegraphics[width=2\columnwidth, clip=on, trim={0 30 0 0}]{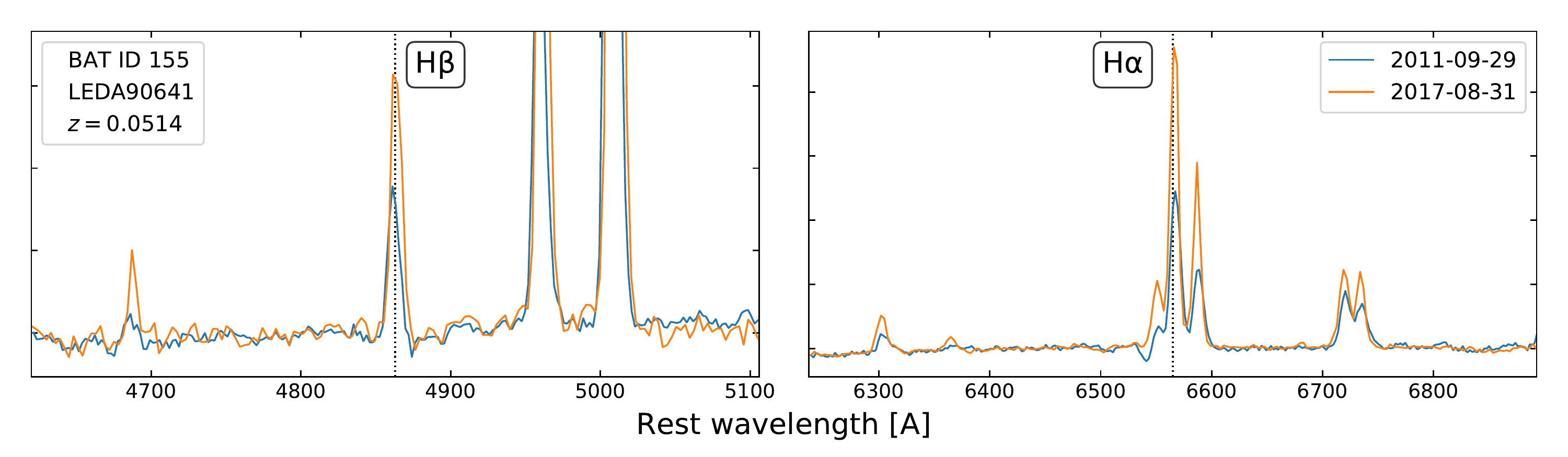}
    \includegraphics[width=2\columnwidth, clip=on, trim={0 30 0 0}]{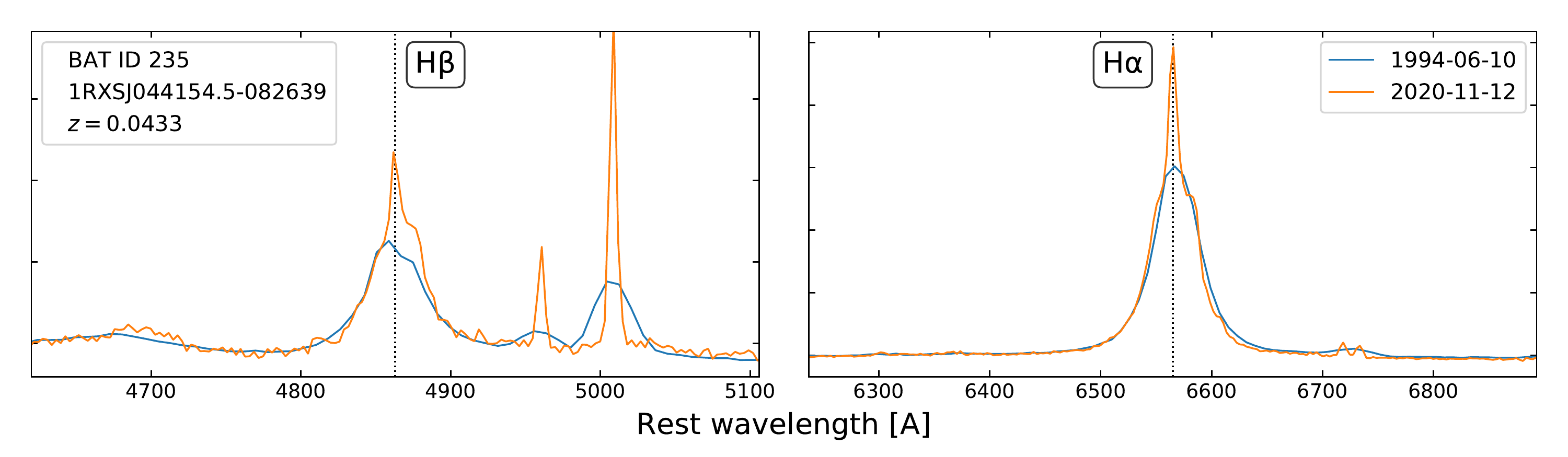}
    \includegraphics[width=2\columnwidth, clip=on, trim={0 0 0 0}]{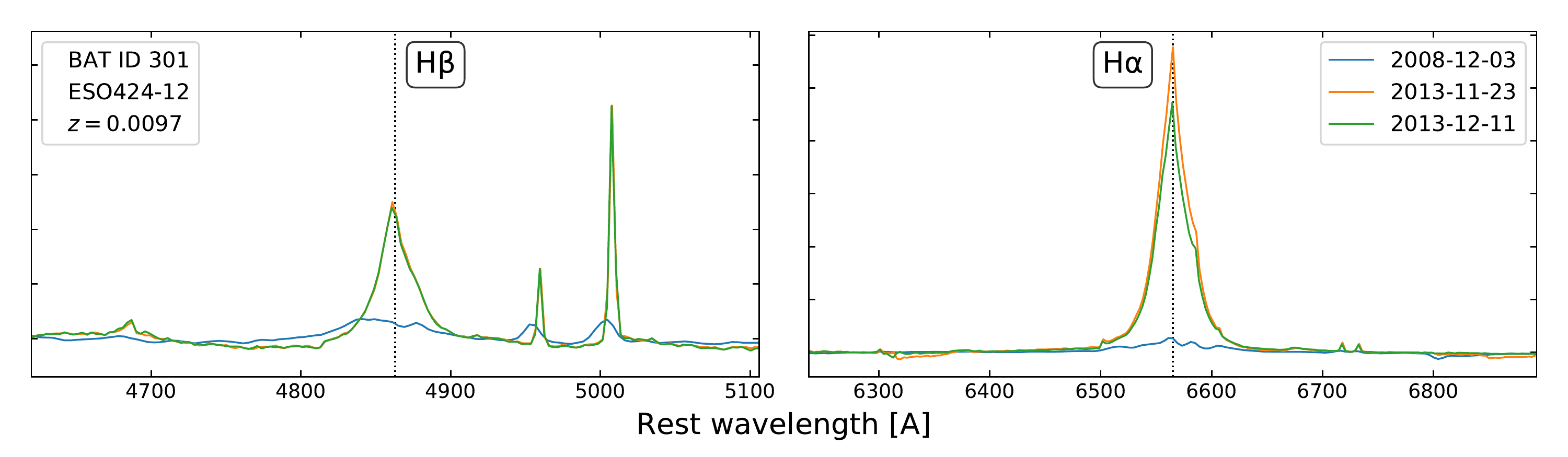}
    \caption{Example flux density spectra for BASS AGN which do not show conclusive evidence for changes in their Balmer emission.
    The majority of our parent sample, like BAT\,150 in the top panel, show no change in their apparent emission line properties.
    Below we show examples where apparent changes in the Balmer line strengths could be explained by aperture effects (BAT\,155), varying spectral resolution (BAT\,235), or a combination of both (BAT\,301). Such objects were also removed as CL AGN candidates during visual inspection. As in Fig.~\ref{fig:specs}, counts have been normalised using the median flux density across each spectral window and dotted lines show the rest-frame wavelengths of Balmer  \hb\ $\lambda$4861 and  \ha\ $\lambda$6563. }
    \label{fig:spec_no_change}
\end{figure*}
\begin{figure*}
    \centering
    \includegraphics[width=2\columnwidth, clip=on, trim={0 30 0 0}]{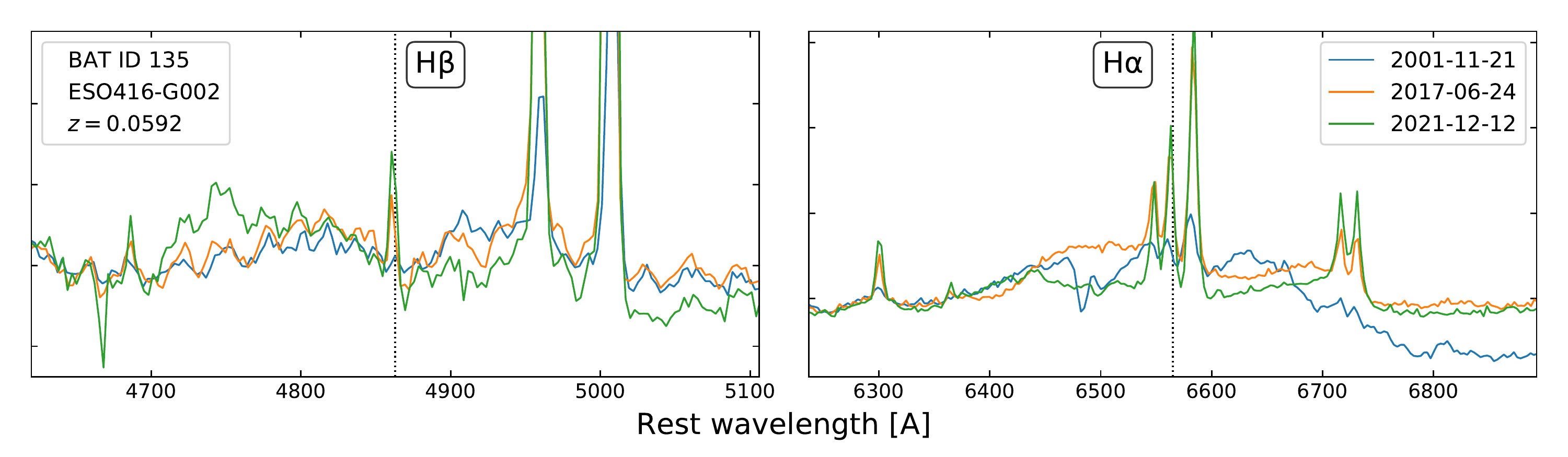}
    \includegraphics[width=2\columnwidth, clip=on, trim={0 30 0 0}]{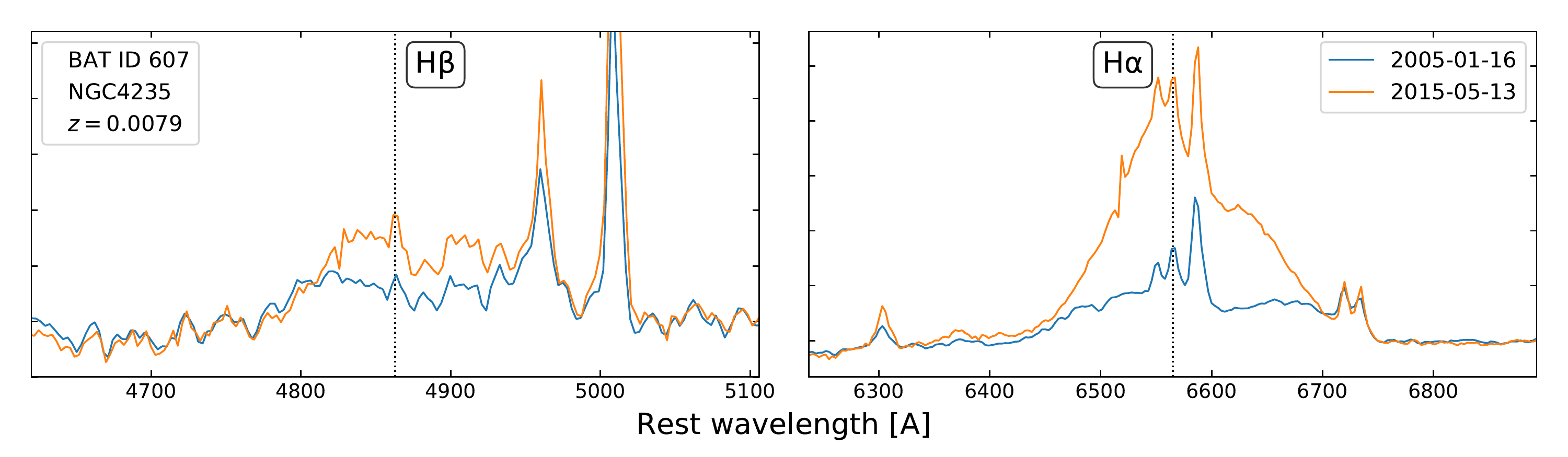}
    \includegraphics[width=2\columnwidth, clip=on, trim={0 0 0 0}]{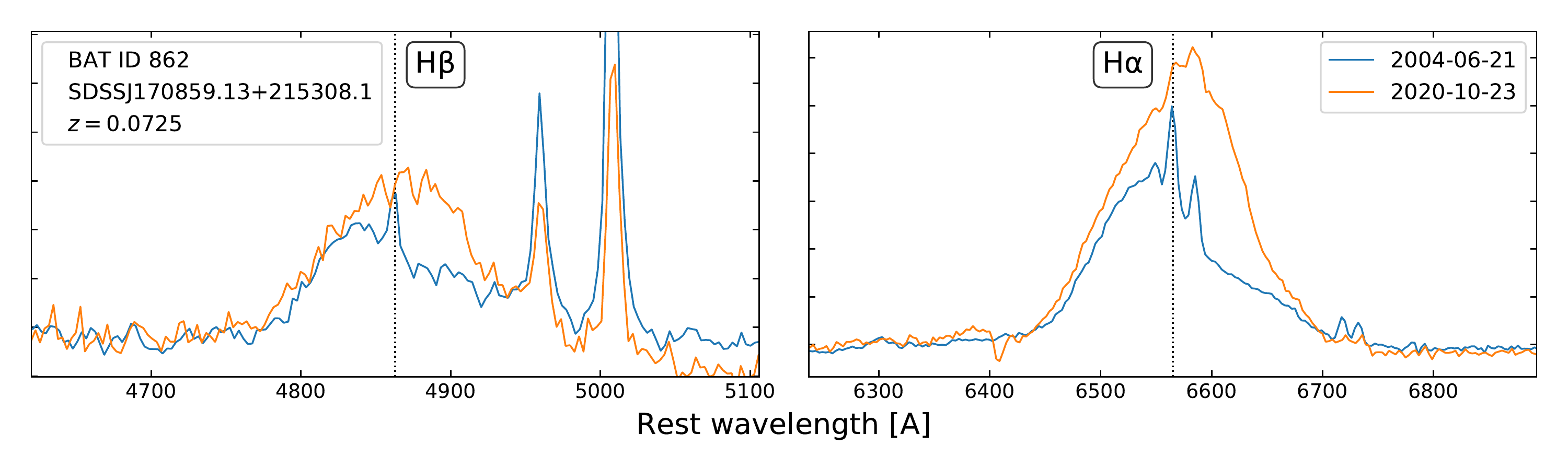}
    \caption{Flux density spectra for three BASS AGN which show evidence for changing Balmer emission morphologies, but which are not included in our final CL sample. BAT\,135 appears to show dramatic change between 2001 and 2017, but the 2001 November spectrum is from the 6dF Galaxy Survey and has somewhat uncertain flux calibration \citep{2004MNRAS.355..747J}, especially around  \ha\ in the red end of the spectrograph. The 2017 June and 2021 December spectra from BASS both show evidence for a very broad, albeit weak, base to the \ha\ emission.
    BAT\,607 and BAT\,862  show changes in the strength and velocity structure of their Balmer emission, but broad \ha\ and \hb\ are present in all epochs and so these objects do not meet our CL criteria for a complete appearance or disappearance of broad line emission.
    As in Fig.~\ref{fig:specs}, counts have been normalised using the median flux density across each spectral window and dotted lines show the rest-frame wavelengths of Balmer  \hb\ $\lambda$4861 and  \ha\ $\lambda$6563. }
    \label{fig:spec_weak_change}
\end{figure*}

In this appendix we discuss examples of BASS AGN from our parent sample with multiple epochs of spectroscopy which were not identified as CL AGN in our visual inspection process. The instruments used to obtain these spectra are given in Table~\ref{tab:nonCS_AGN}.

The vast majority of our parent sample are consistent with no intrinsic changes in their spectral features.
Some small fraction of these show apparent changes in their emission line strengths which could be due to different instrumental resolutions, aperture sizes, or flux calibration procedures. We show three of the clearest examples of such effects in  Fig.~\ref{fig:spec_no_change}.
There are also a small number of objects where the strength of the broad Balmer emission line appears to change, and we show three examples in Fig.~\ref{fig:spec_weak_change}. As discussed in the main text, we only include objects which show the complete appearance or disappearance of a broad line in our CL AGN sample, where it is much more likely that the observed change in the line properties is not due to instrumental effects.

\begin{table}
	\centering
	\caption{Summary of the BASS spectra for the seven objects discussed in Appendix~\ref{sec:app}.}
	\label{tab:nonCS_AGN}
	\begin{tabular}{llcr} % four columns, alignment for each
		\hline
		BAT ID & Counterpart Name & Observation dates & Instruments\\
		% & Counterpart Name
		\hline
		BAT\,135 & ESO\,416\,--\,G\,002
		  & 2001-11-21 & AAO/6dF \\
		&& 2017-06-24 & VLT/Xshooter \\
		&& 2021-12-12 & Magellan/MagE \\
		\hline
		BAT\,150 & 2MASX\,J02502722+4647295
		 & 2011-06-01 & Perkins/DeVeny\\
		&& 2018-09-09 & Palomar/DBSP \\
		&& 2019-08-29 & Palomar/DBSP \\
		\hline
		BAT\,155 & LEDA\,90641
		 & 2011-09-29 &  SPM/BC \\
		&& 2017-08-31 &  Palomar/DBSP \\
		\hline
		BAT\,235 & 1RXS\,J044154.5--082639
		 & 1994-06-10 &  MPG/EFOSC2 \\
		&& 2020-11-12 &  Palomar/DBSP \\
		\hline
		BAT\,301 & ESO\,424\,--\,12
		 & 2008-12-03 &  SPM/BC \\
		&& 2013-11-23 &  VLT/Xshooter \\
		&& 2013-12-11 &  VLT/Xshooter \\
		\hline
		BAT\,607 &  NGC\,4235
		 & 2005-01-16 &  APO/SDSS \\
		&& 2015-05-13 &  VLT/Xshooter \\
		\hline
		BAT\,862 & SDSS\,J170859.13+215308.1
	  	 & 2004-06-21 & APO/SDSS \\
		&& 2020-10-23 & Palomar/DBSP \\
		\hline
	\end{tabular}
\end{table}

%%%%%%%%%%%%%%%%%%%%%%%%%%%%%%%%%%%%%%%%%%%%%%%%%%
% Don't change these lines
\bsp	% typesetting comment
\label{lastpage}
\end{document}